\documentclass{article}

\usepackage{microtype}
\usepackage{graphicx}
\usepackage{subfigure}
\usepackage{booktabs} 

\usepackage{graphicx}

\usepackage{amsmath}
\usepackage{amssymb}

\usepackage{verbatim}

\usepackage{hyperref}

\usepackage{enumitem}
\usepackage{multirow}
\usepackage{longtable}

\usepackage[accepted]{icml2021}

\newcommand{\R}{\mathbb{R}}
\newcommand{\E}{{\mathbb E}}

\icmltitlerunning{Scalable Evaluation of Multi-Agent Reinforcement Learning with Melting Pot}

\begin{document}

\twocolumn[
\icmltitle{Scalable Evaluation of Multi-Agent Reinforcement Learning with Melting Pot}



\icmlsetsymbol{equal}{*}

\begin{icmlauthorlist}
\icmlauthor{Joel Z. Leibo}{equal,deep}
\icmlauthor{Edgar Du{\'e}{\~n}ez-Guzm{\'a}n}{equal,deep}
\icmlauthor{Alexander Sasha Vezhnevets}{equal,deep}
\icmlauthor{John P. Agapiou}{equal,deep}
\icmlauthor{Peter Sunehag}{deep}
\icmlauthor{Raphael Koster}{deep}
\icmlauthor{Jayd Matyas}{deep}
\icmlauthor{Charles Beattie}{deep}
\icmlauthor{Igor Mordatch}{goog}
\icmlauthor{Thore Graepel}{deep}
\end{icmlauthorlist}

\icmlaffiliation{deep}{DeepMind}
\icmlaffiliation{goog}{Google Brain}

\icmlcorrespondingauthor{Joel Z. Leibo}{jzl@google.com}

\icmlkeywords{Machine Learning, ICML}

\vskip 0.3in
]



\printAffiliationsAndNotice{\icmlEqualContribution} 

\begin{abstract}
Existing evaluation suites for multi-agent reinforcement learning (MARL) do not assess generalization to novel situations as their primary objective (unlike supervised-learning benchmarks). Our contribution, Melting Pot, is a MARL evaluation suite that fills this gap, and uses reinforcement learning to reduce the human labor required to create novel test scenarios. This works because one agent's behavior constitutes (part of) another agent's environment. To demonstrate scalability, we have created over 80 unique test scenarios covering a broad range of research topics such as social dilemmas, reciprocity, resource sharing, and task partitioning. We apply these test scenarios to standard MARL training algorithms, and demonstrate how Melting Pot reveals weaknesses not apparent from training performance alone.
\end{abstract}


\section{Introduction}

No broadly accepted benchmark test set for multi-agent reinforcement learning (MARL) research yet exists. This lack of a standardized evaluation protocol has impeded progress in the field by making it difficult to obtain like-for-like comparisons between algorithms. The situation in MARL is now in stark contrast to the status quo in single-agent reinforcement learning (SARL) where a diverse set of benchmarks suitable for different purposes are available (e.g.~\citet{brockman2016openai, fortunato2019generalization, machado2018revisiting, osband2019behaviour, tassa2018deepmind, torrado2018deep}). Further afield, the comparison to the evaluation landscape in other machine learning subfields is even more unfavorable. Datasets like ImageNet~\cite{deng2009imagenet} and their associated evaluation methodology achieve a level of rigor and community acceptance unparalleled in reinforcement learning. 

Within the SARL research community there have been several recent calls to import the methodological stance concerning the primacy of generalization from supervised learning \cite{cobbe2019quantifying, farebrother2018generalization, juliani2019obstacle, zhang2018dissection, zhang2018natural}. However, MARL research is still lagging behind. No one has yet attempted to build a benchmark with the explicit aim of pushing standards for judging multi-agent reinforcement learning research toward making generalization the first-and-foremost goal.

Supervised learning research benefits immensely from having a clear experimental protocol and set of benchmarks that explicitly measure how well methods generalize outside the data to which they were exposed in training \cite{chollet2019measure, deng2009imagenet, lecun2015deep}. This facilitates clear like-for-like comparison between methods, channeling competition between research groups, and driving progress. One problem that arises when trying to import these ideas to reinforcement learning however is that generating a test set of environments is a lot more labor intensive than labeling a set of images. The engineering challenge of creating just a single test environment is akin to designing and implementing a new computer game. Thus calls to appreciate the primacy of generalization in SARL appear sometimes to justify a Sisyphean struggle to create ever more clever and more diverse intelligence tests. 

This obstacle to scalability turns out to be much less severe for research aimed at multi-agent intelligence. In fact, multi-agent approaches have a natural advantage over single-player approaches in the measurement arena. In multi-agent systems, agents naturally pose tasks to one another. Any change to the policy of one agent changes the environment experienced by a set of interdependent others. For instance, if a focal agent learns an effective policy against a fixed set of co-players, it could be rendered useless if the co-players change. This aspect of multi-agent learning is more commonly associated with proposals for ``training time'' ideas like autocurricula \cite{baker2019emergent, bansal2017emergent, leibo2019autocurricula, sukhbaatar2017intrinsic} and open-endedness \cite{clune2019ai}. Here however, we propose to make a different use of it. We can take advantage of multi-agent interaction to create large and diverse sets of generalization tests by pre-training ``background populations'' of agents to use for subsequent evaluation \emph{only}, never training on them---much like the test images in the ImageNet challenge for supervised learning.

Our proposal, Melting Pot, consists of an evaluation methodology and a suite of specific test environments. Its essence is embodied in its central ``equation'':

    \textbf{Substrate} + \textbf{Background Population}  = \textbf{Scenario}

A \emph{scenario} is a multi-agent environment that we use only for testing; we do not allow agents to train in it. The term \emph{substrate} refers to the physical part of the world, it includes: the layout of the map, where the objects are, how they can move, the rules of physics, etc. The term \emph{background population} refers to the part of the simulation that is imbued with agency---excluding the \emph{focal population} of agents being tested.

The Melting Pot research protocol aims to assess and compare multi-agent reinforcement learning algorithms. It is only concerned with test-time evaluation, and so is mostly agnostic to training method.  That is, training-time access to each test's substrate is allowed but we do not mandate how to use it. The suite consists of a collection of zero-shot---i.e. not allowing for test-time learning---test scenarios that preserve a familiar substrate while substituting a new and unfamiliar background population.

Our intention is for Melting Pot to cover the full breadth of different types of strategic situations commonly studied in multi-agent reinforcement learning\footnote{The largest category of extant research that we left unrepresented is communication/language (e.g.~\citet{lazaridou2020emergent, lowe2019pitfalls, mordatch2018emergence}). We see no reason why scenarios engaging these ideas could not be added in the future. Turn-based games (e.g.~\citet{lanctot2019openspiel}) and games involving physics (e.g.~\citet{liu2018emergent}) were also omitted.
}. As such, we have included purely competitive games, games of pure common interest, team-based competitive games, and a range of different kinds of mixed-motivation games including prisoner's dilemma-like social dilemmas and games that stress coordination. The numbers of simultaneous players in each game range from two to sixteen and most substrates have around eight.

Finally, we provide benchmark results on Melting Pot for several different MARL models. Intriguingly, we find that maximizing collective reward often produces policies that are less robust to novel social situations than the policies obtained by maximizing individual reward.

\section{What does Melting Pot evaluate?}

We use the term \emph{multi-agent population learning algorithm} (MAPLA) to refer to any training process that produces a decentralized population of agents capable of simultaneous interaction with one another. Melting Pot evaluates MAPLAs on their fulfillment of three desiderata. They are best introduced by way of an example. Consider the following problem faced by a manufacturer of self-driving cars. The goal is to build a population of agents that will act simultaneously in the world as decentralized individuals. They do not necessarily know in advance whether their cars will be a small minority of the overall number of vehicles on the road, interacting with large numbers of human drivers and self-driving cars from competing manufacturers, or whether their product might prove so popular that it rapidly becomes a majority of the cars on the road. This fraction may even change dynamically from day to day (consider: a competitor might recall their product, or human driving could become illegal). The self-driving fleet (a multi-agent population) must then satisfy the following:

Individual agents and sets of agents sampled from the population must:
\begin{enumerate}
    \item perform well across a range of social situations where individuals are interdependent,
    \item generalize to interact effectively with unfamiliar individuals not seen during training (who may well be human), and
    \item pass a universalization test: answering positively to the question ``what if everyone behaved like that?''. 
\end{enumerate}

The class of MAPLA algorithms is very broad. Most multi-agent reinforcement learning approaches can be made to produce populations. For instance self-play schemes like those used for AlphaGo~\cite{silver2016mastering, silver2017mastering}, AlphaZero~\cite{silver2018general}, FTW (Capture the Flag)~\cite{jaderberg2019human}, hide and seek~\cite{baker2019emergent}, and AlphaStar~\cite{vinyals2019grandmaster} fit in the class of MAPLAs, as does recent work on DOTA~\cite{berner2019dota} and MOBA~\cite{ye2020towards} games, as well as algorithms like MADDPG~\cite{lowe2017multi}, LOLA~\cite{foerster2018learning}, $\text{PSRO}$~\cite{lanctot2017unified}, $\text{PSRO}_{rN}$~\cite{balduzzi2019open}, and Malthusian reinforcement learning~\cite{leibo2019malthusian}.

\section{Related work}

The idea to use agents to create tasks for other agents appears in research on competitive games, such as Capture the Flag~\cite{jaderberg2019human}, DOTA~\cite{berner2019dota} and StarCraft II~\cite{vinyals2019grandmaster}. 
There evaluation against held-out agents was successfully used to measure generalization, and ultimately beat human professionals. 
This idea also appears in contexts where agent interaction is used to drive learning \cite{racaniere2019automated, wang2019paired}. In these cases it was not used to create benchmarks for generalisation.
Zero-shot transfer to new co-players in coordination games was investigated in \citet{hu2020other}.
Several papers~\citet{song2020arena, lowe2017multi} have introduced MARL benchmarks for specific domains, but do not measure generalisation and don't use learning agents to produce evaluation tasks.
Another approach with a long history in game theory involves organizing a competition between strategies submitted by different research groups (e.g.~\citet{Axelrod84}). Doing well in such a competition involves generalization since a submitted agent must play with the other researchers' submissions. \citet{perez2019multi} brought this competition approach to MARL. Differing from all these approaches, Melting Pot covers a broader range of multi-agent interactions and is more explicitly focused on providing a benchmark for generalization.

\section{The Melting Pot protocol}\label{section:protocol_formalism}

\subsection{Definitions}

Our term, \emph{substrate}, refers to a partially observable general-sum Markov game (e.g.~ \citet{shapley1953stochastic, littman1994markov}). In each game state, agents take actions based on a partial observation of the state space and receive an individual reward. The rules of the game are not given to the agents; they must explore to discover them. Thus a Melting Pot substrate is simultaneously a game of \emph{imperfect} information---each player possesses some private information not known to their coplayers (as in card-games like poker)---and \emph{incomplete} information---lacking common knowledge of the rules \citep{harsanyi1967games}. We describe the various substrates available in Melting Pot in Section \ref{substrates}.

Formally, a substrate is an $N$-player partially observable Markov game $\mathcal{M}$ defined on a finite set of states $\mathcal{S}$, observations $\mathcal{X}$, and actions $\mathcal{A}$. The observation function $\mathcal{O} : \mathcal{S} \times \{1, \dots , N\} \rightarrow \mathcal{X}$, specifies each player's view of the state space. In each state, each player $i$ selects an individual action $a_i \in \mathcal{A}$. Following their joint action $\mathbf{a} = (a_1, \dots , a_N) \in \mathcal{A}^N$, the state change obeys the stochastic transition function  $\mathcal{T} : \mathcal{S} \times \mathcal{A}^N \rightarrow \Delta(\mathcal{S})$, where $\Delta(\mathcal{S})$ denotes the set of discrete probability distributions over $\mathcal{S}$. After a transition, each player receives an individual reward defined by $\mathcal{R}: \mathcal{S} \times \mathcal{A}^N \times \{1, \dots , N\} \rightarrow \R$.

A \emph{policy} $\pi: \mathcal{X} \times \mathcal{A} \times \mathcal{X} \times \mathcal{A} \times \dots \times \mathcal{X} \rightarrow \Delta(\mathcal{A})$ is a probability distribution over a single agent's actions, conditioned on that agent's history of observations and previous actions. Policies are not transferable between substrates, since $\mathcal{X}$ and $\mathcal{A}$ can differ between them. Let $\pi \in \Pi_{\mathcal{M}}$ indicate a policy defined on the substrate $\mathcal{M}$, and consider a \emph{joint policy} formed by a tuple of individual policies $\boldsymbol{\pi} = (\pi_1, \dots, \pi_n) \in {\Pi_{\mathcal{M}}}^n$. We call this (factorized) joint policy \emph{compatible} with the $N$-player substrate $\mathcal{M}$ if $n = N$. A compatible joint policy is necessary to sample episodes from the interaction between the individual policies and the substrate.

Given a compatible joint policy $\boldsymbol{\pi}$ on substrate $\mathcal{M}$, we measure the performance of each individual policy within this context as the \emph{individual return} $R_i(\boldsymbol{\pi}|\mathcal{M})$---the expected total reward for player $i$. We then measure the performance of the joint policy using the \emph{per-capita return}---the mean individual return:

\begin{equation*}
  \bar{R}(\boldsymbol{\pi}|\mathcal{M}) = \frac{1}{N} \sum_{i=1}^N R_i(\boldsymbol{\pi}|\mathcal{M})
\end{equation*}

A \emph{population} for an $N$-player substrate $\mathcal{M}$ is a distribution $f(\Pi_{\mathcal{M}})$ over individual policies. A population for $\mathcal{M}$ can therefore create a compatible joint policy $\boldsymbol{\pi}$ for $\mathcal{M}$ by independently sampling $N$ individual policies from $f$: $\boldsymbol{\pi} \sim f(\pi_1)\dots f(\pi_N)$. Performance on a substrate by a population is measured by the expected per-capita return:
\begin{equation*}
  \bar{R}(f|\mathcal{M}) = \frac{1}{N} \sum_{i=1}^N \E_{\pi_1 \sim f} \dots \E_{\pi_N \sim f} R_i(\boldsymbol{\pi}|\mathcal{M})
\end{equation*}

\subsection{Testing}

Let a \emph{scenario configuration} for an $N$-player substrate $\mathcal{M}$ be a binary vector $\mathbf{c} = (c_1,\dots,c_N) \in \{0, 1\}^N$ of $n$ zeros and $m$ ones that indicates whether each player $i$ is a \emph{focal player} ($c_i = 1$), or a \emph{background player} ($c_i = 0$). Let the \emph{background population} for a substrate $\mathcal{M}$ be a distribution $g(\Pi_{\mathcal{M}})$ that is used to sample individual policies for the background players. Now, from the perspective of the focal players, the $N$-player substrate $\mathcal{M}$ is \emph{reduced} to an equivalent $m$-player substrate $\mathcal{M}^\prime$ (via marginalization of the transition and reward functions by the background player policies). We call this reduced $m$-player substrate ---formed from a substrate, configuration, and background population---a \emph{test scenario}.

Performance on a test scenario by a \emph{focal population} $f$ is measured by the expected per-capita return as before, except that background players are excluded from the mean:
\begin{align*}
  \bar{R}(f|\mathcal{M}, \mathbf{c}, g)
      =& \bar{R}(f|\mathcal{M}^\prime) \\
      =& \frac{1}{m} \sum_{i=1}^N c_i \E_{\pi_1 \sim h_1} \dots \E_{\pi_N \sim h_N} R_i(\boldsymbol{\pi}|\mathcal{M})
\end{align*}
where $h_i(\pi) = f(\pi)^{c_i}g(\pi)^{1-c_i}$.

When focal players outnumber background players, we say the test scenario is in \emph{resident} mode. These scenarios usually test the emergent cooperative structure of the population under evaluation for its robustness to interactions with a minority of unfamiliar individuals not encountered during training. When background players outnumber focal players, we say the test scenario is in \emph{visitor} mode. One common use case for visitor-mode scenarios in Melting Pot is to test whether an individual from the focal population can observe the conventions and norms of the dominant background population and act accordingly (without retraining).

We use another kind of test scenario to test universalization. In this case, we have no background players, but instead of independently sampling each focal policy $\pi_i$ from $f$, we sample from $f$ once and use this policy repeatedly. So the joint policy consists of $N$ copies of the same policy $(\pi, \dots, \pi) \in \Pi^N$ where $\pi \sim f$. We call this the \emph{universalization} mode. It answers the rhetorical question ``how would you like it if everyone behaved like you?''. Such universalization is an important part of human moral psychology \cite{levine2020logic}, at least in some cultures \cite{henrich2020weirdest}.
Because this universalization test scenario does not require a background population, it could easily be incorporated into a training process, in which case it would not be a test of generalization. However, we included it because it is useful in diagnosing failure cases.

\subsection{Training}

During training, MAPLA $F$ is provided with unlimited access to a substrate $\mathcal{M}$, which it uses to train a population $f$. Thus, the whole training process may be represented as $F[\mathcal{M}] \mapsto f(\Pi_{\mathcal{M}})$. The purpose of Melting Pot is to evaluate the MAPLA by measuring the performance of the resulting \emph{focal population} $f$.

We do this by measuring the per-capita return of the focal population when used to sample focal players in our test scenarios. Note that the learning algorithms only have access to the raw substrates and not the background populations. This means that policies sampled from the focal population must show good zero-shot generalization to unseen test scenarios.

Consider a MAPLA where $N$ separate policies are trained together in a $N$-player substrate, and $f$ selects a trained policy uniformly. In order for $f$ to perform well in test scenarios, self-play should have adequately explored similar behavior to that present in the background population.

Our definition of population deliberately penalizes heterogeneous specialization. Since policies must be independently sampled at test-time, the training algorithm cannot control their joint distribution, and so cannot prescribe a fixed division of labor. To perform well on test scenarios, trained agents should be generalists. Division of labor is possible in Melting Pot, but it works best when it self-organizes at test time with individuals taking on their roles in response to the ongoing behavior of others. In the cases where specialization is most important, successful populations should feature significant redundancy in order to be robust enough to do well in Melting Pot.

Melting Pot focuses only on test-time evaluation, and is agnostic to the method of training. For example, during training, the substrate can be augmented to give agents privileged access to the rewards and observations of co-players. This privileged information is not present in test scenarios, so policies that rely on it will generalize poorly. But it can be useful for providing auxiliary targets to improve the training of internal representations (e.g. ``centralized training and decentralized execution''~\cite{kraemer2016multi, lowe2017multi, oliehoek2016concise}).

\subsection{Secondary evaluation metrics}

We propose the \emph{focal-population per-capita return} as a primary evaluation metric, to test the performance of a learning algorithm in a novel social situation. This is because, first and foremost, we want Melting Pot to provide a rigorous and clearly interpretable evaluation metric that highlights unsolved problems and compares innovative algorithms to one another.

However, when evaluating the suitability of trained agents for a practical application, there will be additional considerations, which can be assessed from secondary evaluation metrics using the same test scenarios. For example, impacts on the background population may be an indicator of the impacts the trained agents might have on humans in the real world. So we can measure the \emph{background-population per-capita return} to see if it is negatively impacted by the introduction of the focal population. This could be useful to study whether the joint policy of the focal agents produces negative externalities---``side effects'' that impact the broader population while sparing the focal population, dovetailing well with research on value alignment and AI safety \cite{soares2014aligning, amodei2016concrete}. Or following \citet{perolat2017}, we can measure the \emph{inequality} of the background-population individual returns to see if any benefit or harm arising from having introduced the focal population is fairly shared, perhaps connecting fruitfully to beneficial and cooperative AI research agendas \cite{russell2015research, dafoe2020open}.

\begin{figure}[ht]
		\includegraphics[width=\columnwidth]{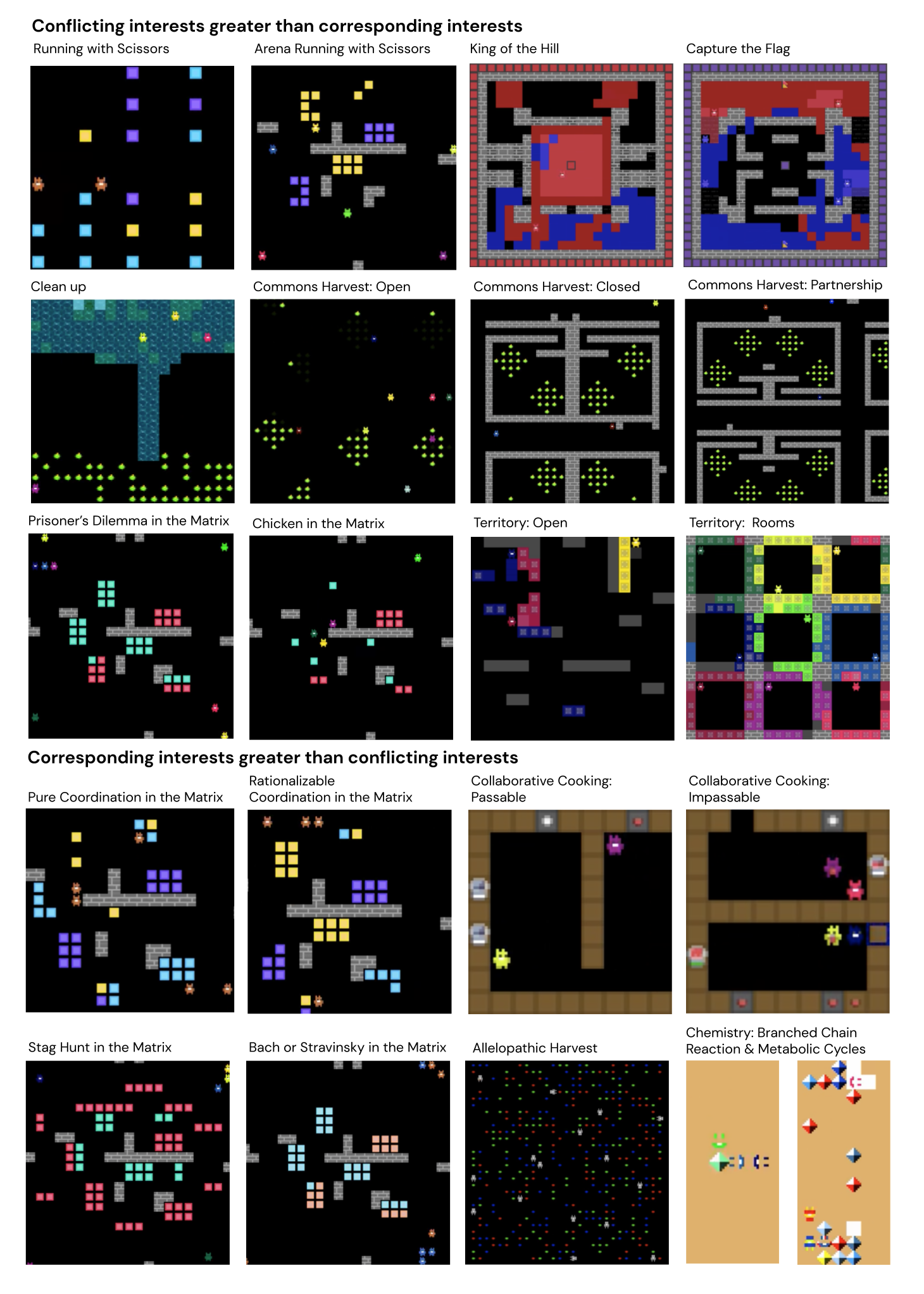}
		\caption{Overview of substrates.}
\end{figure}

\begin{figure*}[ht]
{\centering
		\includegraphics[width=0.95\linewidth]{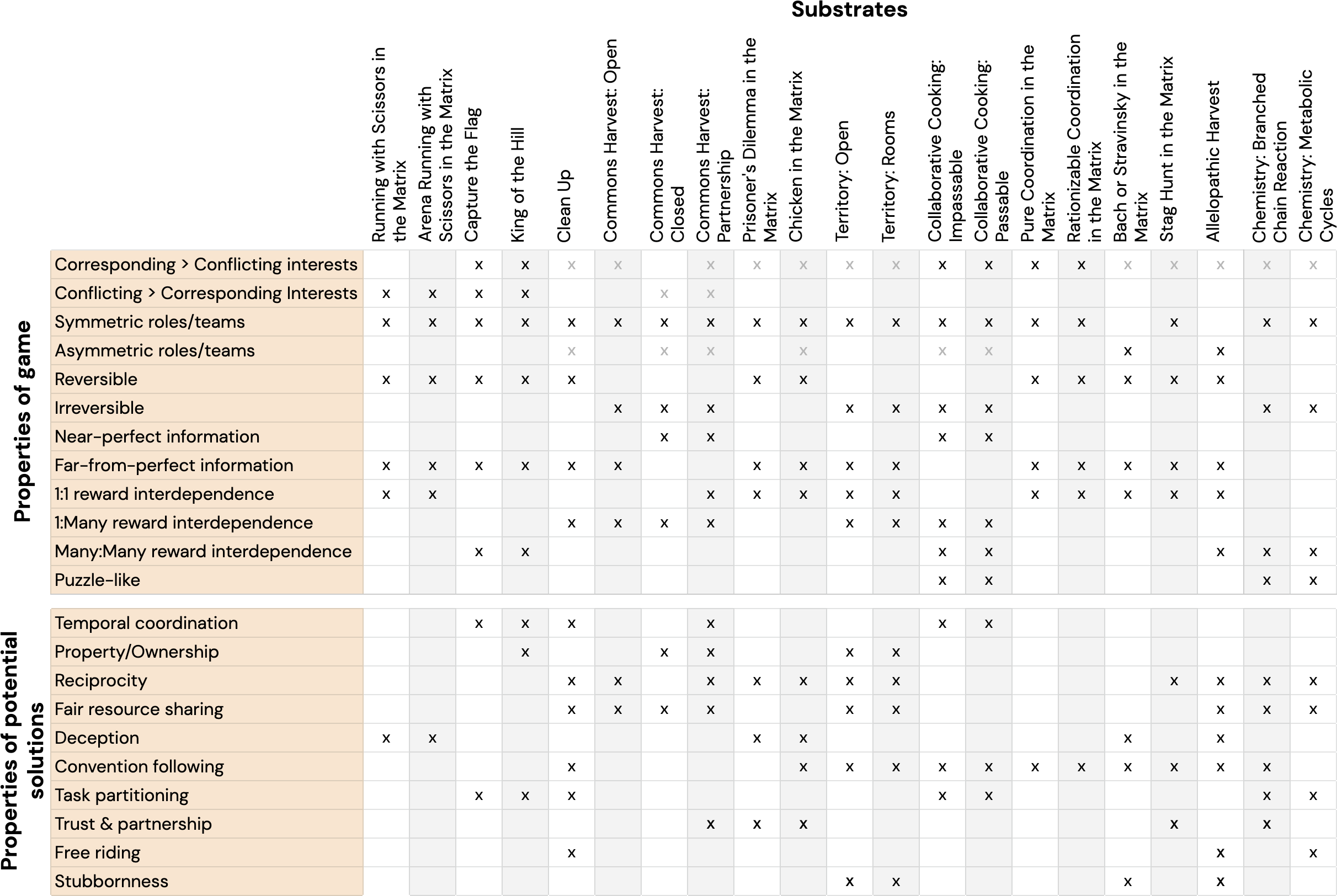}
		\caption{Multi-agent concepts engaged by each substrate. Gray ticks represent secondary characteristics. Properties highlighted here are only intended as a rough guide to the concepts at play in each substrate. They should not be taken as a serious theoretical attempt to classify multi-agent situations.}\label{fig:concepts}}
\end{figure*}

\section{Description of the substrates} \label{substrates}

Fig.~\ref{fig:concepts} provides a rough guide to the strategic and social intelligence concepts covered by the suite. See the appendix for additional details that describe the precise setup for all substrates and test scenarios. Some substrates share common game mechanics. All \emph{* in the Matrix} games all share the same pattern: Players can collect items representing some number of choices (e.g. defect and cooperate or rock, paper, and scissors), and when they encounter each other their inventory is used to dispense rewards according to the payoff matrix of a classic matrix game.

\subsection{Conflicting greater than corresponding interests}

\emph{Running with Scissors in the Matrix} first appeared in \citet{vezhnevets2020options}. Two individuals gather rock, paper, or scissor resources in the environment, and can challenge others to a `rock, paper scissor' game, the outcome of which depends on the resources they collected.  It is possible (though not trivial) to observe the policy that one's partner is starting to implement, and to take countermeasures. This induces a wealth of possible feinting strategies. \emph{Arena Running with Scissors in the Matrix} extends the game to eight players. 

In \emph{Capture the Flag} teams of players can expand their territory by painting the environment, which gives them an advantage in a confrontation with the competing team. The final goal is capturing the opposing team's flag. Payoffs are common to the entire winning team. \emph{King of the Hill} has the same dynamics except the goal is to control the ``hill'' region in the center of the map. For both substrates there are scenarios where agents play with familiar teammates against unfamiliar opponents as well as scenarios where ad-hoc teamwork is needed \cite{stone2010ad}.

\emph{Clean up} is a social dilemma where individuals have to balance  harvesting of berries for reward with cleaning a river that suppresses berry growth if it gets too dirty \cite{hughes2018inequity}. As cleaning the river is a public good, individuals are motivated to harvest instead of clean.

In \emph{Commons Harvest: Open} individuals harvest apples that fail to regrow if a patch is exhausted. Preserving a patch requires all agents to show restraint in not harvesting the last apple \cite{perolat2017}. \emph{Commons Harvest: Closed} has the apples in rooms that can be defended by a single player, alleviating the risk of others over-harvesting. In \emph{Commons Harvest: Partnership} it takes two players to defend a room, requiring effective cooperation both in defending and in not over-harvesting.

\emph{Prisoner's Dilemma in the Matrix} mirrors the classic matrix game that exposes tension between individual and collective reward. In \emph{Chicken in the Matrix}, both players attempting to defect leads to the worst outcome for both. These substrates target similar concepts to the Coins game of \citet{lerer2018maintaining}, though they are somewhat more complex---in part because they have more players (eight versus two).

In \emph{Territory: Open} individuals can claim a territory for reward by coloring it. They can find a peaceful partition, but also have the option of irreversibly destroying potentially rewarding territory rendering it useless for everyone. \emph{Territory: Rooms} has segregated rooms that strongly suggest a partition individuals could adhere to.

\subsection{Corresponding greater than conflicting interests}

\emph {Collaborative Cooking: Impassable} is inspired by \cite{carroll2019utility, wang2020too}'s work on an Overcooked-like environment. Players need to collaborate to follow recipes, but are separated by an impassable kitchen counter so no player can complete the objective alone. In \emph{Collaborative Cooking: Passable}, players can have access to all sections of the kitchen, which allows individual players to sequentially perform all subtasks unilaterally (but less efficiently). 

In \emph{Pure Coordination in the Matrix} all individuals need to converge on the same color choice to gain reward when they encounter each other. Which convention emerges in a given population is entirely arbitrary, and all players are indifferent between the possible conventions. In \emph{Rationalizable Coordination in the Matrix} the choices are of different values, suggesting an optimal color to converge on.

\emph{Bach or Stravinsky in the Matrix} and \emph{Stag Hunt in the Matrix} focus on coordination. In the former, coordination is tied to unfairness and ``stubbornness'' could play a role. In the latter, coordination is associated with risk for the individual. It engages similar concepts to \citet{peysakhovich2017prosocial}'s Markov Stag Hunt game, though it is more complex---in part due to it being an eight player game (instead of two-player).

Combining the above dynamics, in \emph{Allelopathic Harvest}, players can increase the growth-rate of berries by planting berries in the same color. However, for each player, a different berry color is intrinsically more rewarding. This creates tensions between groups of players and a free-rider problem between individuals who prefer to consume rather than plant \cite{koster2020model}.

In \emph{Chemistry}, individuals control chemical reactions by transporting molecules around the space. \emph{Chemistry: Branched Chain Reaction} requires alternation between two specific reactions. Combining molecules efficiently requires coordination, but can also lead to exclusion of players. In \emph{Chemistry: Metabolic cycles}, individuals benefit from two different cyclic reaction networks and must coordinate to keep them both running.

\section{Extending Melting Pot}\label{section:extending}

We want to grow Melting Pot over time and ultimately create a comprehensive platform where most aspects of social intelligence can be assessed. To that end, we designed Melting Pot around the need to establish a scalable process through which it can be expanded. This led us to consider not just modular environment components (which we have), but also a modular process for contributing new scenarios.

A scenario consists of two parts: a substrate, and a background population. We built substrates on DMLab2D \cite{beattie2020deepmind} using an entity-component system approach similar to that of modern game engines like Unity \cite{unity}. Members of the background population are RL agents. We call them \emph{bots} to distinguish them from the agents in the focal population. A Melting Pot substrate emits events when interactions occur between agents, or agents and the environment, such as one player zapping another player or eating an apple. Events can be conditional on the identities of the players involved or the location where the interaction occurred.

\begin{figure}[ht]
		\includegraphics[width=\columnwidth]{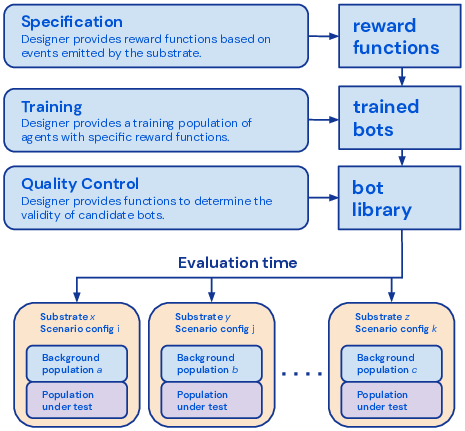}
		\caption{\label{fig:bot_production} The process for extending Melting Pot.}
\end{figure}

Our approach to creating background populations involves three steps: (1) specification, (2) training, and (3) quality control (Fig.~\ref{fig:bot_production}). We describe each in turn.

\emph{1. Specification:} The designer typically starts with an idea of what they want the final bot's behavior to look like. Since substrate events provide privileged information about other agents and the substrate, we can often easily specify reward functions that induce the right behaviour.
This is a much easier task than what focal agents need to solve---learning from only pixels and the final reward. 
However, sometimes the desired behavior is difficult to specify using a single reward function. In these cases, we generate background populations using techniques inspired by hierarchical reinforcement learning~\cite{sutton2011horde,schaul2015universal,sutton1999between}; in particular reward shaping~\cite{sutton2018reinforcement} and ``option keyboard''~\cite{barreto2019option}.
We create a basic portfolio of behaviors by training bots that use different environment events as the reward signal  (as in Horde~\cite{sutton2011horde}), and then chain them using simple Python code.
This allows us to express complex behaviours in a ``if this event, run that behaviour'' way. For example, in \emph{Clean Up} we created a bot that only cleans if other players are cleaning.
These bots had a special network architecture based on FuN~\cite{vezhnevets2017feudal}, with goals specified externally via substrate events rather than being produced inside the agent. See appendix for details.

\emph{2. Training:} The decision at this stage is how to train the background population. The thing to keep in mind is that the bots must generalize to the focal population. To this end, we chose at least some bots---typically not used in the final scenario---that are likely to develop behaviors resembling that of the focal agent at test time. For instance, in \emph{Running With Scissors in the Matrix}, we train rock, paper, and scissors specialist bots alongside ``free'' bots that experience the true substrate reward function. 

\emph{3. Quality control:} Bot quality control is done by running 10--30 episodes where candidate bots interact with other fixed bots. These other bots are typically a mixture of familiar and unfamiliar bots (that trained together or separately). We verify that agents trained to optimize for a certain event, indeed do. We reject agents that fail to do so. 

\begin{figure*}[ht]
		\includegraphics[width=\linewidth]{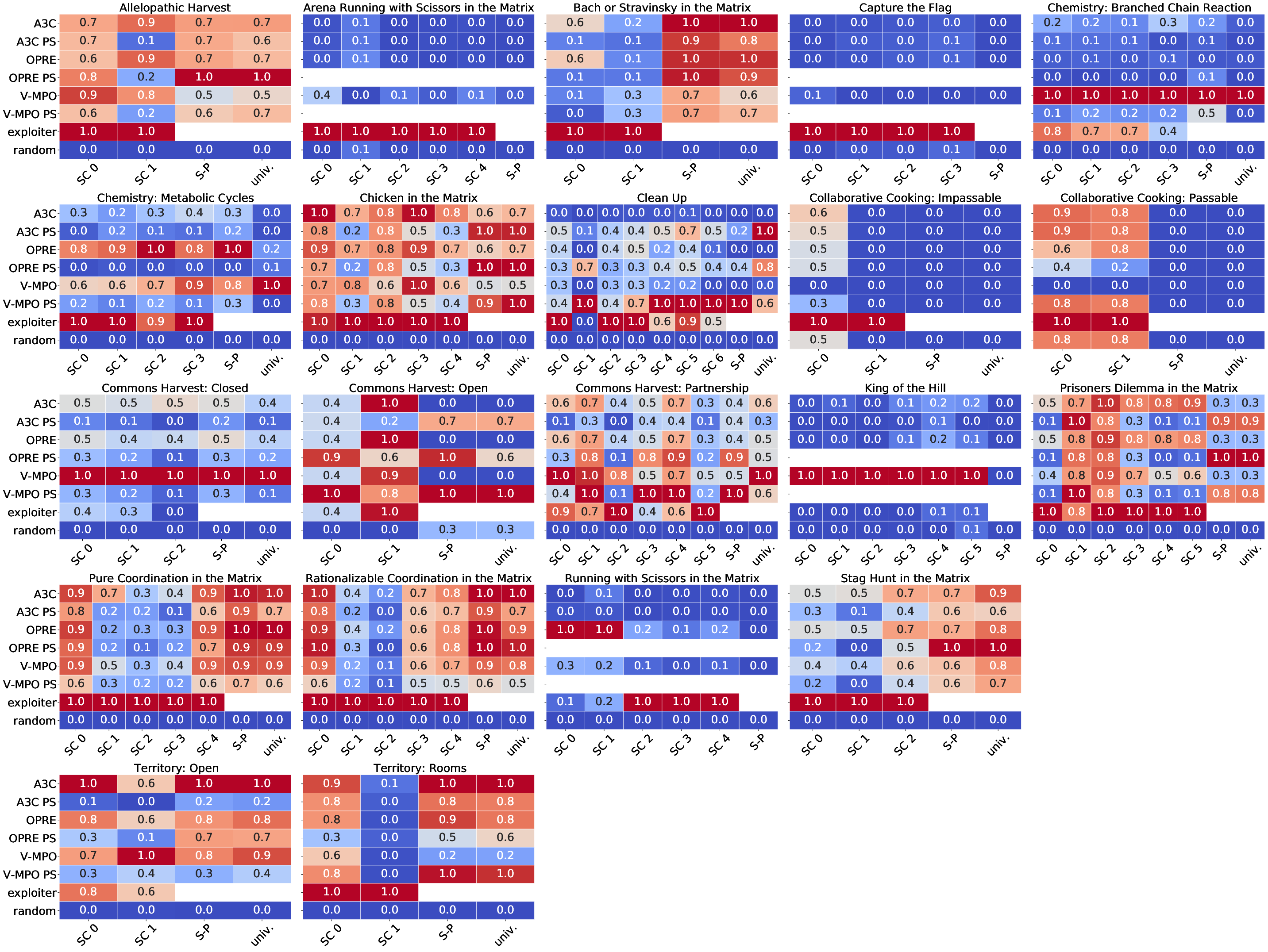}
		\caption{MAPLA performance scores on test scenarios for each substrate. Rows correspond to the agent training algorithms, columns to scenarios. PS stands for prosocial, SC for scenario, S-P for self-play score (i.e. the training performance), univ. for universalization scenario. Note: we did not train prosocial variants on zero-sum substrates since the per-capita return is zero by definition.}
		\label{fig:agents_vs_scenario}
\end{figure*}

\begin{figure*}[ht]
		\includegraphics[width=\linewidth]{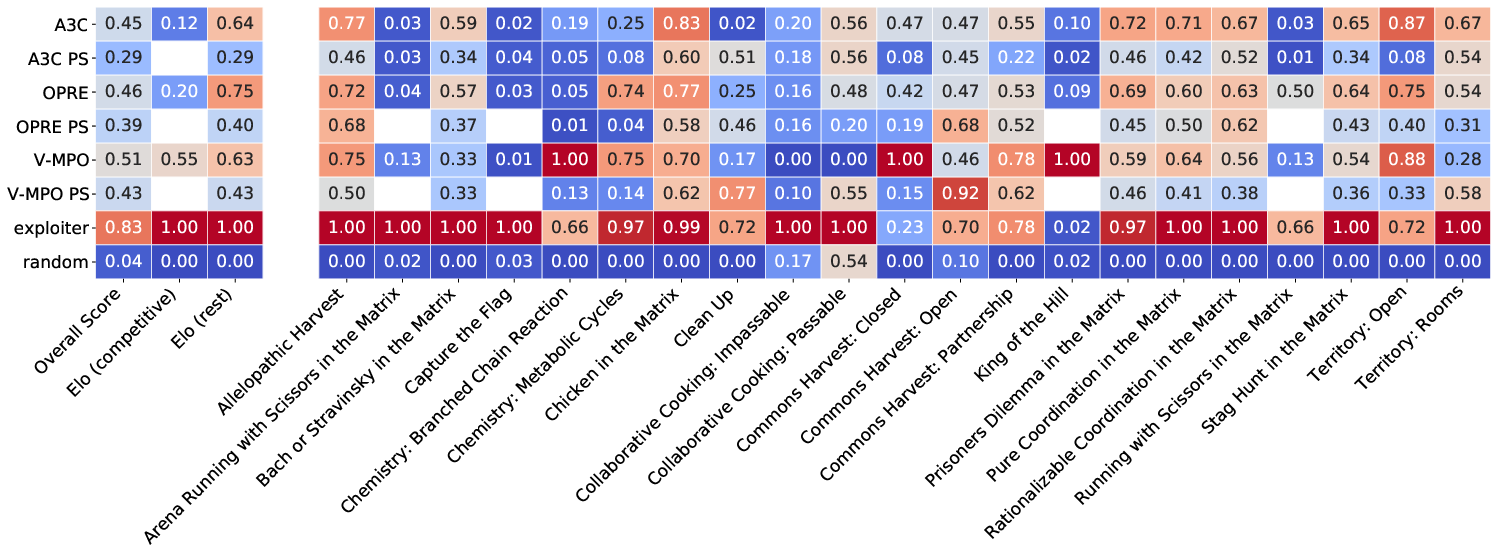}
		\caption{MAPLA performance scores averaged over each substrate's scenarios. Three columns on the left show: average score per algorithm; Elo per agent on competitive substrates, and Elo on the rest the suite. Elo is min-max normalised such that $0$ corresponds to lowest performing agent, and $1$ to the highest performing one.}
		\label{fig:agents_vs_substrate_test}
\end{figure*}

\section{Experiments}

To demonstrate the use of Melting Pot, we provide benchmark MARL results for a number of agent architectures.

For each agent architecture, we performed 21 training runs---one for each substrate. Within each training run, we trained a group of $N$ agents---one for each player in the $N$-player substrate. Every agent participated in every training episode, with each agent playing as exactly one player (selected randomly on each episode). Each agent was trained for $10^9$ steps. At test time, we set the focal population to be the uniform distribution over the $N$ agents.

The different architectures we trained are: A3C~\cite{mnih2016asynchronous}, V-MPO~\cite{Song2020VMPO}, and OPRE~\cite{vezhnevets2020options}. 
A3C is a well established, off-the-shelf RL method. V-MPO is relatively new and state-of-the-art on many single agent RL benchmarks. OPRE was specifically designed for MARL. We also trained \textit{prosocial} variants of all three algorithms, which directly optimized the per-capita return (rather than individual return), by sharing reward between players during training. Optimizing for collective return as a surrogate objective has previously been used for collaborative games (substrates) and social dilemmas \cite{claus1998dynamics,peysakhovich2017prosocial}, and our experiments here allow us to investigate whether it generalizes well.

All agent architectures had the same size convolutional net and LSTM. The OPRE agent had additional hierarchical structure in its policy as described in~\cite{vezhnevets2020options}. V-MPO had a pop-art layer~\cite{hessel2019multi} for normalizing the value function. A3C minimized a contrastive predictive coding loss \cite{oord2018representation} as an auxiliary objective \cite{jaderberg2016reinforcement} to promote discrimination between nearby time points via LSTM state representations (a standard augmentation in recent work with A3C). See the appendix for additional implementation details.

We also use two special agents: ``exploiters'' and ``random''. 
Each exploiter is an A3C agent trained directly on a single test scenario, using their individual return as the reward signal without further augmentation. Exploiters trained for up to $10^9$ steps. The random agent selects actions uniformly at random, and ignores input observations. Together, exploiters and the random agent provide a rough estimate of the upper and lower bounds (respectively) of performance on the test scenarios.
To contextualize the range of agent returns, we min-max normalize the focal per-capita returns to get a \emph{performance score} that is between 0 (for the worst agent) and 1 (for the best agent). Often the score of 0 corresponds to the random agent and 1 to the exploiter. However, for some scenarios this is not the case. We discuss the reasons below.

Fig.~\ref{fig:agents_vs_scenario} presents the full results, showing the score obtained by each agent on each test scenario. 
Fig.~\ref{fig:agents_vs_substrate_test} presents an overview, showing the average test scenario score for each substrate. Alongside the total average score, we present Elo scores~\cite{balduzzi2018re,hunter2004mm}, which are computed separately for competitive substrates and the rest.

On average, the top performing agent was V-MPO~\cite{Song2020VMPO}, followed by OPRE~\cite{vezhnevets2020options}, and A3C~\cite{mnih2016asynchronous}. 
All three performed similarly in mixed-motivation games, but V-MPO outperforms in competitive games like \emph{King of the Hill}. However, ranking agents' by skill is complicated~\cite{balduzzi2018re}, and dependent on the contribution of each test scenario to the overall score. Here we did only a simple evaluation which is far from the only option.

There is scope to dramatically improve performance on the most challenging substrates. The \textit{Collaborative Cooking} substrates proved impossible for these agents to learn, with no agent obtaining a self-play (training) score above random. In \textit{Arena Running with Scissors} and \textit{Capture The Flag}, agents are far below exploiter performance on the test scenarios.

Exploiters do not always achieve the highest score. In some cases (\emph{Running with Scissors in the Matrix}, \emph{King of the Hill}), exploiters fail because the scenario bots are strong opponents and learning to exploit them is hard. Here, the MAPLAs have the advantage of a gentler curriculum: in self-play training, all opponents are of a similar skill level, which ramps up as they all learn.
In other cases (\emph{Commons Harvest Open}, \emph{Clean Up}), exploiters fail due to their selfish reward maximization in games where cooperation is required. In these cases, a better performance upper bound might be obtained by exploiters with within-group reward sharing.

Agents often exhibit a degree of overfitting. In Fig.~\ref{fig:agents_vs_scenario} we see agents with high self-play scores (obtained during training), but low test-scenario scores.
For example, in \emph{Bach or Stravinsky in the Matrix} and \emph{Stag Hunt in the Matrix}, prosocial OPRE and prosocial A3C achieve a similar self-play score to regular OPRE and A3C, but obtain a closer to random score in the test scenarios. 
This overfitting is due to prosocial agents only learning cooperation strategies during self-play training, and so becoming exploitable by defectors at test time.

Overall, prosocial agents underperformed their selfish counterparts, but the picture is nuanced.
Optimizing for per-capita return can be difficult because it complicates credit assignment, and creates spurious reward ``lazy agent'' problems \cite{sunehag2018value, rashid2018qmix}. However, in the social dilemma \emph{Clean Up}, only prosocial agent architectures managed to learn policies that were significantly better than random. This suggests that doing well on Melting Pot will require agents to be able to contingently balance selfishness and prosociality.

The universalization scenarios can diagnose issues with the of division of labour.
For example, in \emph{Chemistry: Metabolic Cycles}, OPRE performs well in self-play and other scenarios, but has low universalization scores. This means that some agents in the population learned specialized policies that expect other agents to behave in a particular way. Although such division of labour can create efficiency, it also makes populations less robust.

\section{Conclusion}

Here we have presented Melting Pot: an evaluation suite for MAPLAs that evaluates generalization to novel social situations. Melting Pot engages with concepts that have long been neglected by research in artificial intelligence. Solutions to the problems posed here seem to require agents that understand trust, generosity, and forgiveness, as well as reciprocity, stubbornness, and deception.

Melting Pot is scalable. We have used reinforcement learning to reduce human labor in environment design. This is how we rapidly created the diverse set of $\sim 85$ scenarios considered so far. Since Melting Pot has been openly released, it can be extended by any interested researchers. In addition, since the effectiveness of the bots in test scenarios is itself advanced by improvements in the performance of learning systems, Melting Pot will likewise improve over time by reincorporating the latest agent technology into new background populations and test scenarios.

The Melting Pot open-source project is available at\\ \url{https://github.com/deepmind/meltingpot}.

\newpage

\appendix

\part*{Appendix}

\newlist{SC}{enumerate}{1}
\setlist[SC]{label=SC~\arabic*, start=0, align=left, labelwidth=2em, leftmargin=\labelwidth+\labelsep}

\section*{Contents}

\begin{enumerate}[wide=0pt, widest=99,leftmargin=3em, labelsep=*, label=\textbf{\Alph*.}]\setcounter{enumi}{0}
    \item {\bf Secondary evaluation metrics}
    
    \item {\bf Substrate details}
    
    \item {\bf Agent architecture details}

    \item {\bf Training setup}

    \item {\bf Scenario details}
    
    \item {\bf Raw performance scores}
    
    \item {\bf Acknowledgements}
    
    \item {\bf References}
\end{enumerate}

\setcounter{section}{0}
\renewcommand*{\theHsection}{chX.\the\value{section}}

\counterwithin{figure}{section}

\section{Secondary evaluation metrics}


In the article we discussed additional metrics of interest when evaluating the impact of the focal population on the background population. Here we demonstrate two such secondary evaluation metrics: \emph{background per-capita return}, and \emph{background positive-income equality}.

\begin{figure}[b]
        \includegraphics[width=\columnwidth]{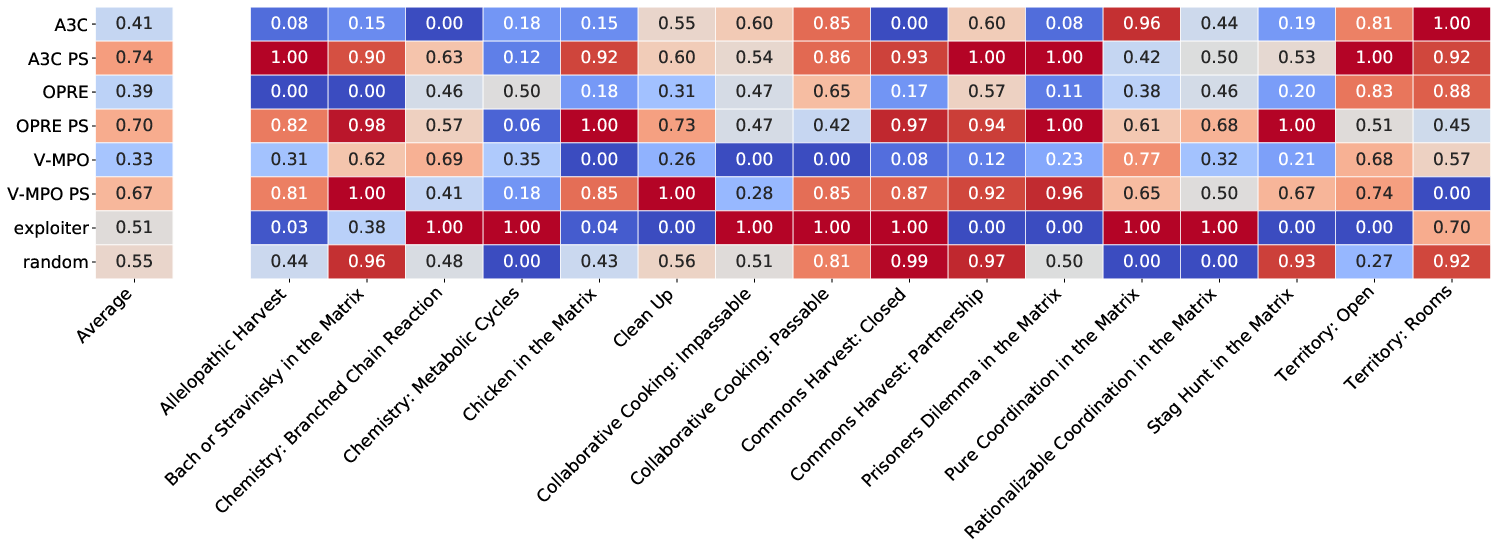}
        \caption{Background per-capita return (normalized).}
        \label{fig:background_per_capita_return}
\end{figure}

\emph{Background per-capita return} is the average return of players drawn from the background population. We min-max normalize the raw return into a score as we did for the focal per-capita return.

\emph{Background positive-income equality} \cite{perolat2017} is a measure of how uniformly returns are distributed between agents in the background population. It is given by the complement of the Gini coefficient of the background-player positive returns:
\begin{align*}
  Q(\mathbf{r}) = 1 - \frac{
          \sum_{i=1}^m \sum_{j=1}^m \left| r^+_i - r^+_j \right|
      }{
          2 m \sum_{i=1}^m r^+_i
      }
\end{align*}
where $m$ is the number of background players, and $r^+_i = \max(0, r_i)$ is the positive part of the return $r_i$ to background player $i$. Background positive-income equality is 0 when only one background player receives a positive return, and 1 when all background players receive the same return.

\begin{figure}[b]
        \includegraphics[width=\columnwidth]{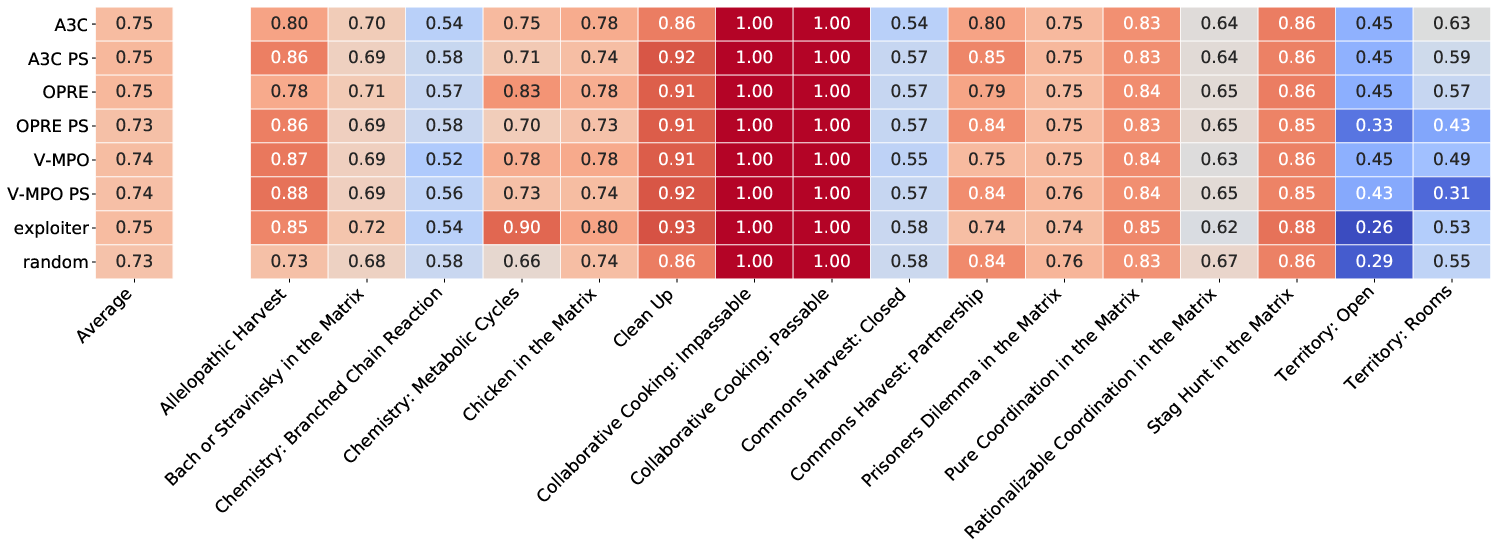}
        \caption{Background positive-income equality.}
        \label{fig:background_positive_income_equality}
\end{figure}

Fig.~\ref{fig:background_per_capita_return} and Fig.~\ref{fig:background_positive_income_equality} show how these secondary metrics are impacted by different focal populations in different substrates. We omitted the zero sum (pure competition) environments from these analyses since the impact of the focal population on the background population is negative by definition. To ensure each scenario had a large enough sample size of background players, we also excluded ``resident-mode'' test scenarios where focal players outnumbered background players.

On average, the per-capita return of the background population is higher when interacting with the prosocial variant of each agent (Fig.~\ref{fig:background_per_capita_return}). Since the prosocial agents were explicitly trained to maximize the reward of other players, it is unsurprising that they would most benefit the background population. Notable exceptions to this are the \textit{Chemistry} and \textit{Collaborative Cooking} substrates where the prosocial variants had poor training and universalization performance (indicating ``lazy agent'' issues c.f.~\cite{sunehag2018value, rashid2018qmix}). Since the prosocial variants did not learn the required behavior for these highly collaborative substrates they also did not make good partners.

In addition, prosocial variants usually did not have any negative impact on background positive-income equality (Fig.~\ref{fig:background_positive_income_equality}), indicating no downside to the benefit imparted to the background-population. A notable exception here is \textit{Chicken in the Matrix} where prosocial variants (vastly) improved background per-capita return, but at the cost of (mildly) increased inequality. This was likely a consequence of prosocial agents increasing the number of cooperators (a benefit to the background population), but not every background-player can find a prosocial player to partner with/exploit.

\newpage
\section{Substrate details}

\subsection{Common to all substrates}

Unless otherwise stated, all substrates have the following common rules:
\begin{itemize}
    \item Episodes last 1000 steps.
    \item Sprites are $8 \times 8$ pixels.
    \item The agents have a partial observability window of $11 \times 11$ sprites, offset so they see more in front than behind them. The agent sees 9 rows in front of itself, 1 row behind, and 5 columns to either side.
    \item Thus in RGB pixels, the size of each observation is $88 \times 88 \times 3$. All agent architectures used here have RGB pixel representations as their input. 
    \item Movement actions are: forward, backward, strafe left, strafe right, turn left, and turn right.
\end{itemize}

\subsection{Shared by multiple substrates}

\subsubsection{* in the Matrix}

This mechanism was first described in \cite{vezhnevets2020options}.

Agents can move around the map and collect resources of $K$ discrete types. In addition to movement, the agents have an action to fire the interaction beam. All agents carry an inventory with the count of resources picked up since last respawn. The inventory is represented by a vector 
\[ \rho = \left( \rho_1, \dots, \rho_K \right)\text{. }
\]
Agents can observe their own inventory but not the inventories of their coplayers. When another agent is zapped with the interaction beam, an interaction occurs. The resolution of the interactions is driven by a traditional matrix game, where there is a payoff matrix $A$ describing the reward produced by the pure strategies available to the two players. The resources map one-to-one to the pure strategies of the matrix game. Unless stated otherwise, for the purposes of resolving the interaction, the zapping agent is considered the row player, and the zapped agent the column player. The actual strategy played depends on the resources picked up before the interaction. The more resources of a given type an agent picks up, the more committed the agent becomes to the pure strategy corresponding to that resource. In particular, an agent with inventory $\rho$ plays the mixed strategy with weights
\[ v = \left( v_1, \dots, v_K \right)\]
where
\[ v_i = \frac{\rho_i}{\sum_{j=1}^K \rho_j} \text{. } \]

The rewards $r_\text{row}$ and $r_\text{col}$ for the (zapping) row and the (zapped) column player, respectively, are assigned via
\begin{align*}
    r_{\text{row}} &= v_{\text{row}}^T \, A_\text{row} \, v_{\text{col}}\\
    r_{\text{col}} &= v_{\text{row}}^T \, A_{\text{col}} \, v_{\text{col}}
\end{align*}

If the game is symmetric then $A_{\text{row}} = A_\text{col}^T$.

To obtain high rewards, an agent could either collect resources to ensure playing a Nash strategy in the matrix game, or correctly identify what resource its interaction partner is collecting and collect the resources that constitute a best response. Most substrates have eight simultaneous players, so individuals must also decide who from the group to interact with.

Unless stated otherwise, after an interaction the player with the smaller reward is considered to have ``lost'' the interaction and gets removed from the game for 200 steps. Since episodes last 1000 steps, there are within each episode typically four ``rounds'' when all agents respawn around the same time and try to collect resources and interact with one another. Players experience varying numbers of interactions per ``round'' since it depends whether or not they ``win'' them (losing players disappear and must wait till the next round to interact again). Ties are resolved in favour of the zapping agent (removing the zapped agent). The inventory of the losing agent in the interaction is reset to the initial value (by default, all zeros). The winner's inventory is not reset.

\subsubsection{Chemistry}

\begin{figure}[t]
    	\includegraphics[width=\linewidth]{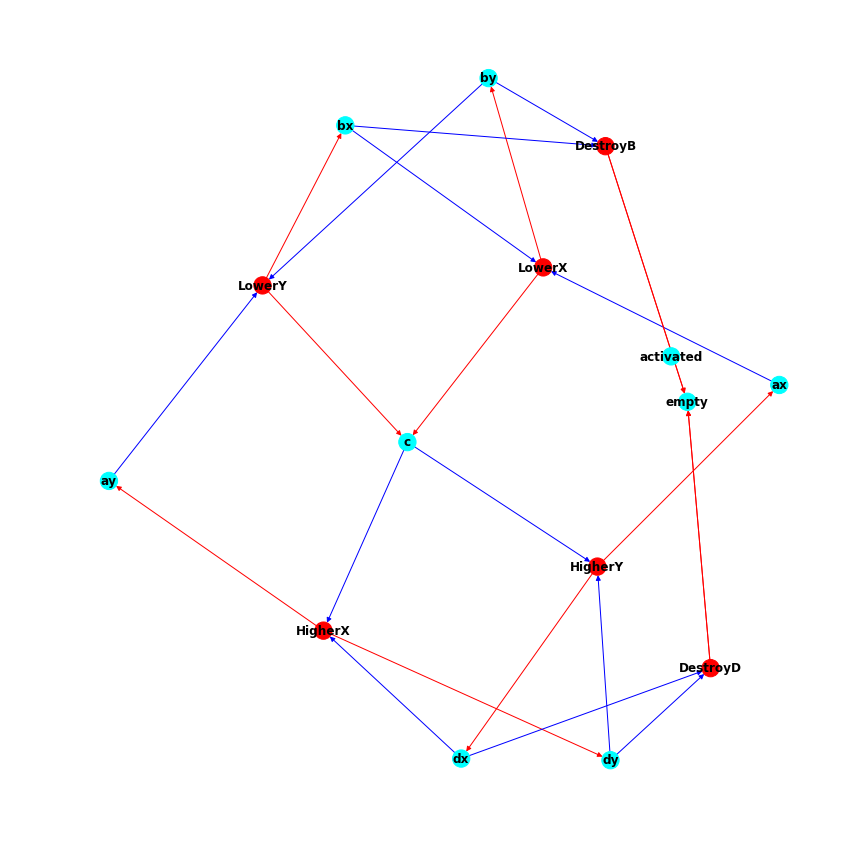}
    	\caption{Reaction Graph for molecules in Chemistry: Branched Chain Reaction}\label{fig:chemistry_branched}
\end{figure}

\begin{figure}[t]
    	\includegraphics[width=\linewidth]{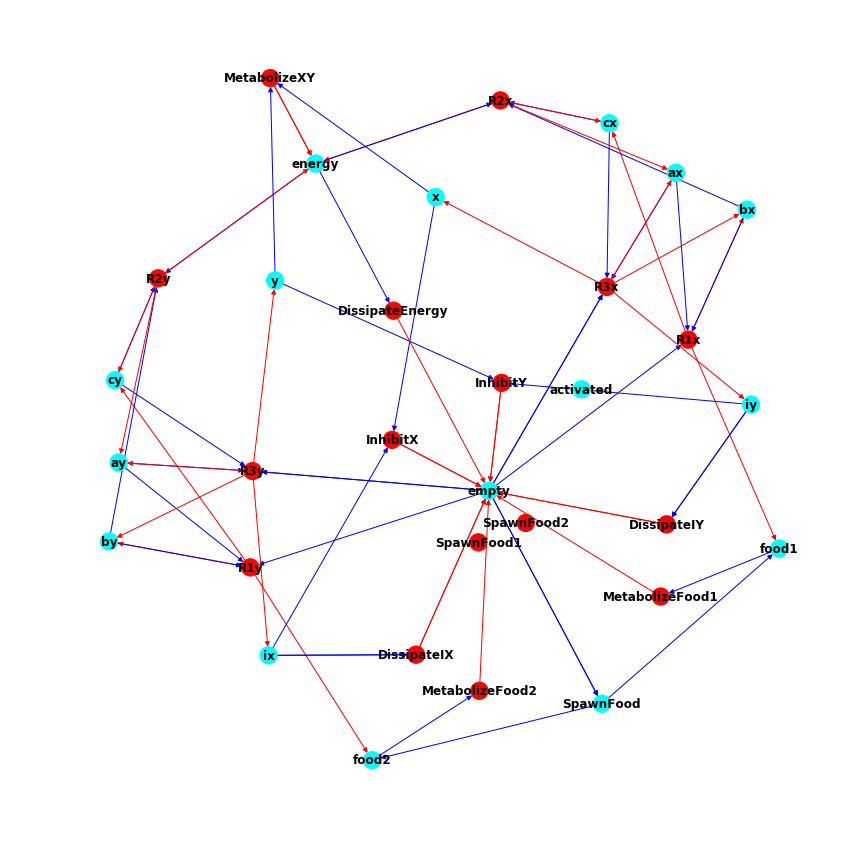}
    	\caption{Reaction Graph for molecules in Chemistry: Metabolic Cycles}
        \label{fig:chemistry_metabolic}
\end{figure}

Reactions are defined by a graph (see Fig.~\ref{fig:chemistry_branched}, Fig.~\ref{fig:chemistry_metabolic}), which together with a map setting initial molecules defines a substrate, occur stochastically when reactants are brought near one another. Agents can carry a single molecule around the map with them at a time. Agents are rewarded when a specific reaction---such as metabolizing food---occurs with the molecule in their inventory participating as either a reactant or a product. Each reaction has a number of reactants and products, occurs at different rates that can depend on if it is in an agent's inventory or not. As an example, metabolizing food in the metabolic cycles substrate has a much higher rate in the inventory where it generates a reward than outside where it represents the food molecule dissipating.

\subsubsection{Commons Harvest}

This mechanism was first described in \citet{janssen2010lab} and adapted for multi-agent reinforcement earning by \citet{perolat2017}.

Apples are spread around and can be consumed for a reward of $1$. Apples that have been consumed regrow with a per-step probability that depends on the number of current apples in a $L^2$ norm neighborhood of radius $2$. The apple regrowth probabilities are $0.025$ when there are three or more apples in the neighborhood, $0.005$ if there are exactly two apples, $0.001$ if there is exactly one, and $0$ if there are no apples in the neighborhood. As a consequence, a patch of apples that collectively doesn't have other apples within a distance of $2$ from them, can be irrevocably lost if all apples in the patch are consumed. Therefore, agents must exercise restraint when consuming apples within a patch. Notice that in a single agent situation, there is no incentive to collect the last apple in a patch (except near the end of the episode). However, in a multi-agent situation, there is an incentive for any agent to consume the last apple rather than risk another agent consuming it. This creates a \emph{tragedy of the commons} from which the substrate derives its name.

\subsubsection{The painting mechanic in King of the Hill and Capture the Flag}

There is a red team and blue team. Players  can ``paint'' the ground anywhere by using their zapping beam. If they stand on their own color then they gain health up to a maximum of 3 (so they are more likely to win shootouts). They lose health down to 1 from their default of 2 when standing on the opposing team's color (so they are more likely to lose shootouts in that case). Health recovers stochastically, at a fixed rate of $0.05$ per frame. It cannot exceed its maximum, determined by the color of the ground the agent is standing on.

Players also cannot move over their opposing team's color. If the opposing team paints the square underneath their feet then they get stuck in place until they use their own zapping beam to re-paint the square underneath and in front of themselves to break free. In practice this slows them down by one frame (which may be critical if they are being chased).

Friendly fire is impossible; agents cannot zap their teammates.

\subsubsection{The claiming mechanic in Territory: Open and Territory: Rooms}

Players cannot walk through resources, they are like walls.

Players can claim a resource in two ways: (1) by touching it, and (2) by using a ``claiming beam'', different from the zapping beam, which they also have. Claimed resources are colored in the unique color of the player that claimed them. Unclaimed resources are gray.

Once a resource has been claimed a countdown begins. After 100 timesteps, the claimed resource becomes active. This is visualized by a white and gray plus sign appearing on top. Active resources provide reward stochastically to the player that claimed them at a rate of 0.01 per timestep. Thus the more resources a player claims and can hold until they become active, the more reward they obtain.

The claiming beam is of length 2. It can color through a resource to simultaneously color a second resource on the other side. If two players stand on opposite sides of a wall of resources of width 2 and one player claims all the way across to the other side (closer to the other player than themselves) then the player on the other side might reasonably perceive that as a somewhat aggressive action. Less aggressive of course than the other option both players have: using their zapping beam. If any resource is zapped twice then it gets permanently destroyed. It no longer functions as a wall or a resource, allowing players to pass through.

Like resources, when players are hit by a zapping beam they also get removed from the game and never regenerate. This is different from other substrates where being hit by a zapping beam is not permanent removal. In territory once a player has been zapped out it is gone. All resources it claimed are immediately returned to the unclaimed state.

\subsection{Details of specific substrates}

\textbf{Allelopathic Harvest}\footnote{For a video of \textit{Allelopathic Harvest}, see \url{https://youtu.be/ESugMMdKLxI}.}
Sixteen players can increase the growth-rate of berries by planting berries in the same color. However, each player has a particular berry color that is intrinsically more rewarding. This creates tensions between groups of players and a free-rider problem between individuals who prefer to consume rather than plant \cite{koster2020model}.

Individuals are rewarded for consuming ripe berries. They get a reward of 2 for consuming a red berry, and a reward of 1 for consuming a berry of any other color. At the start of each episode, berries are initialized to be unripe and evenly distributed over the three colors. Individuals can replant unripe berries to any color. Thus each individual experiences a tension between their incentive to immediately consume ripe berries and their incentive to plant unripe---and red colored---berries.

The environment (a 29x30 plane) is filled with 348 berry plants of 3 colors (reset at the start of each episode; 116 per color).

Berry ripening depends stochastically on the number of berries sharing the same color that have been planted.  Initially all berries are in an unripe state. Each berry has a probability $p$ to ripen on each step, dependent on the number $b$ of berries of the same color across the whole map; $p = 5\times 10^{-6}b$ (the `allelopathic' mechanic (inspired by \citep{leibo2019malthusian}). Investing in establishing one color throughout the map ``a monoculture'' is prudent because it can be done relatively rapidly (if all players join in) and by doing so, all players will be able to harvest berries at a much faster rate for the remainder of the episode.

Players can move around in the environment and interact with berries in several ways. Players can use a planting beam to change the color of unripe berries to one of the other berry-type colors (the 'harvest' mechanic). Players can also walk over ripe berries to consume them. Ripe berries cannot be replanted, and unripe berries cannot be consumed. Players' avatars are recolored after using their planting beam to the same color they turned the berry into. Players' avatars are also stochastically recolored to a fixed white color when they eat a berry (probability inversely proportional to the highest berry fraction). These rules have the effect that past eating/planting actions often remains visible for others until a new action is taken. 

Each player also has the ability to zap other agents with a beam. It could be used either as a punishment mechanism or as a way to compete over berries. Getting zapped once freezes the zapped player for 25 steps and applies a visible mark to the player indicating that they have been punished. If a second zap is received within 50 steps, the player is removed for 25 steps and receives a penalty of $-10$ reward. If no zap is received for 50 steps, the mark fades. After each use of the zapping beam it is necessary to wait through a cooldown period of 4 steps before it can be used again.

Episodes last 2000 steps. The action space consists of movement, rotation, use of the 3 planting beams, and use of the zapping beam (10 actions total).

\textbf{Arena Running with Scissors in the Matrix}\footnote{For a video of \textit{Arena Running with Scissors in the Matrix}, see \url{https://youtu.be/esXPyGBIf2Y}.}
Same dynamics as \textit{Running with Scissors} but with eight players. The environment is not conducive to forming alliances since rewards are gained in each pairwise conflict. Initial values for the inventory are $(1, 1, 1)$ instead of zeros. The matrix for the interaction is:
\[ A_\text{row} = A_\text{col}^T =
\begin{bmatrix}
\hphantom{+}0 & -1 & +1\\
+1 & \hphantom{+}0 & -1\\
-1 & +1 & \hphantom{+}0
\end{bmatrix}
\text{ .} \]

\textbf{Bach or Stravinsky in the Matrix}\footnote{For a video of \textit{Bach or Stravinsky in the Matrix}, see \url{https://youtu.be/SiFjSyCp2Ss}.}
Individuals collect resources that represent `Bach' or `Stravinsky' and compare inventories in an encounter. Consequences of this inventory comparison are congruent with the classic Bach or Stravinsky matrix game. This game exposes a tension between reward for the group and fairness between individuals. Unlike other * in the matrix games, Bach or Stravinsky is asymmetric. At the start of each episode, half the players (blue avatars) are assigned to always be the row player in all their interactions, and the other half (orange avatars) are assigned to always be the column player. There is no effect when two players with the same row/column assignment try to interact with one another. The game only resolves when row and column players interact with one another. The winner's inventory is also reset after an interaction. Because this game is asymmetric, there is a different matrix $A_\text{row}$ for the row player and $A_\text{col}$ for the column player. The matrices for the interaction are:
\[ A_{\text{row}} =
\begin{bmatrix}
3 & 0\\
0 & 2
\end{bmatrix}
\text{, } \]
and
\[ A_{\text{col}} =
\begin{bmatrix}
2 & 0\\
0 & 3
\end{bmatrix}
\text{ .} \]

\textbf{Capture the Flag}\footnote{For a video of \textit{Capture the Flag}, see \url{https://youtu.be/VRNt55-0IqE}.}
Teams of players can expand their territory by painting the environment, which gives them an advantage in a confrontation with the competing team. The final goal is capturing the opposing team's flag. Payoffs are common to the entire winning team. Indicator tiles around the edge of the map and in its very center display which teams have their own flag on their base, allowing them the possibility of capturing their opponent's flag by bringing it to their own base/flag. When indicator tiles are red then only the red team can score. When indicator tiles are blue then only the blue team can score. When the indicator tiles are purple then both teams have the possibility of scoring (though neither is close to doing so) since both flags are in their respective home bases.

\textbf{Chemistry: Branched Chain Reaction}\footnote{For a video of \textit{Chemistry: Branched Chain Reaction}, see \url{https://youtu.be/ZhRB-_ruoH8}.}
Individuals are rewarded by driving chemical reactions involving molecules. They need to suitably coordinate the alternation of branches while keeping certain elements apart that would otherwise react unfavourably, so as not to run out of molecules required for continuing the chain. Combining molecules efficiently requires coordination but can also lead to exclusion of players.

\textbf{Chemistry: Metabolic cycles}\footnote{For a video of \textit{Chemistry: Metabolic cycles}, see \url{https://youtu.be/oFK9VujhpeI}.}
Individuals benefit from two different food generating cycles of reactions that both rely on energy that dissipates. Bringing together side products from both cycles generates new energy such that the cycles can continue. The population needs to keep both cycles running. 

\textbf{Chicken in the Matrix}\footnote{For a video of \textit{Chicken in the Matrix}, see \url{https://youtu.be/uhAb2busSDY}.}
Individuals can gather resources of different colors. Players' encounters are resolved with the same payout matrix as the game 'Chicken', in which both players attempting to take advantage of the other leads to the worst outcome for both. Collecting red resources pushes one's strategy choice toward playing `hawk` while collecting green resources pushes it toward playing `dove`. The matrix for the interaction is:
\[ A_\text{row} = A_\text{col}^T =
\begin{bmatrix}
3 & 2\\
5 & 0
\end{bmatrix}
\text{ .} \]

\textbf{Clean Up}\footnote{For a video of \textit{Clean Up}, see \url{https://youtu.be/jOeIunFtTS0}.}
Clean Up is a seven player game. Players are rewarded for collecting apples (reward $+1$). In \textit{Clean Up}, apples grow in an orchard at a rate inversely related to the cleanliness of a nearby river. The river accumulates pollution at a constant rate. Beyond a certain threshold of pollution, the apple growth rate in the orchard drops to zero. Players have an additional action allowing them to clean a small amount of pollution from the river. However, the cleaning action only works on pollution within a small distance in front of the agents, requiring them to physically leave the apple orchard to clean the river. Thus, players maintain a public good of orchard regrowth through effortful contributions. Players are also able to zap others with a beam that removes any player hit from the game for 50 steps \cite{hughes2018inequity}.

A group can achieve continuous apple growth in the orchard by keeping the pollution levels of the river consistently low over time. However, on short timescales, each player would prefer to collect apples in the orchard while other players provide the public good of keeping the river clean. This creates a tension between the short-term individual incentive to maximize reward by staying in the orchard and the long-term group interest of a clean river

\textbf{Collaborative Cooking: Impassable}\footnote{For a video of \textit{Collaborative Cooking: Impassable}, see \url{https://youtu.be/yn1uSNymQ_U}.}
Inspired by \citet{carroll2019utility, wang2020too}'s work on an \textit{Overcooked}-like environment. Players need to collaborate to follow recipes. They are separated by an impassable kitchen counter, so no player can complete the objective alone. Observation window is $5 \times 5$.

\textbf{Collaborative Cooking: Passable}\footnote{For a video of \textit{Collaborative Cooking: Passable}, see \url{https://youtu.be/R_TBitc3hto}.}
Same as \textit{Collaborative Cooking: Impassable} except players can pass each other in the kitchen, allowing less coordinated yet inefficient strategies by individual players. Observation window is $5 \times 5$.

\textbf{Commons Harvest: Closed}\footnote{For a video of \textit{Commons Harvest: Closed}, see \url{https://youtu.be/ZHjrlTft98M}.}
Same as \textit{Commons Harvest: Open} except it has rooms full of apples that can be defended by a single player, alleviating the risk of others over-harvesting a patch of apples. Individuals can defend a region from invasion, effectively converting the difficult multi-agent problem into a set of independent single agent problems, each of which can be solved much more easily \cite{perolat2017}.

\textbf{Commons Harvest: Open}\footnote{For a video of \textit{Commons Harvest: Open}, see \url{https://youtu.be/ZwQaUj8GS6U}.}
Sixteen player game. Individuals harvest apples, that fail to regrow if a patch of apples is exhausted. Preserving a patch would require all agents to show restraint in not harvesting the last apple \cite{perolat2017}.

\textbf{Commons Harvest: Partnership}\footnote{For a video of \textit{Commons Harvest: Partnership}, see \url{https://youtu.be/ODgPnxC7yYA}.}
Same as \textit{Commons Harvest: Closed} except that it takes two players to defend a room of apples, requiring effective cooperation in defending and not over-harvesting a patch of apples. It can be seen as a test that agents can learn to trust their partners to (a) defend their shared territory from invasion, and (b) act sustainably with regard to their shared resources. This is the kind of trust born of mutual self interest. To be successful, agents must recognize the alignment of their interests with those of their partner and act accordingly.

\textbf{King of the Hill}\footnote{For a video of \textit{King of the Hill}, see \url{https://youtu.be/DmO2uqGBPco}.}
Same painting and zapping dynamics as \textit{Capture the Flag} except the goal is to control the ``hill'' region in the center of the map. The hill is considered to be controlled by a team if at least 80\% of it has been colored in that team's color. The status of the hill is indicated by indicator tiles around the map an in the center. Red indicator tiles mean the red team is in control. Blue indicator tiles mean the blue team is in control. Purple indicator tiles mean no team is in control.

\textbf{Prisoner's Dilemma in the Matrix}\footnote{For a video of \textit{Prisoner's Dilemma in the Matrix}, see \url{https://youtu.be/bQkEKc1zNuE}.}
Eight individuals collect resources that represent `defect' (red) or `cooperate' (green) and compare inventories in an encounter. Consequences of the inventory comparison are congruent with the classic \textit{Prisoner's Dilemma} matrix game. This game exposes tension between reward for the group and reward for the individual. The matrix for the interaction is
\[ A_\text{row} = A_\text{col}^T  =
\begin{bmatrix}
3 & 0\\
4 & 1
\end{bmatrix}
\text{ .} \]

\textbf{Pure Coordination in the Matrix}\footnote{For a video of \textit{Pure Coordination in the Matrix}, see \url{https://youtu.be/5G9M7rGI68I}.}
Players---who in this case cannot be identified as individuals since they all look the same---can gather resources of three different colors. All eight individuals need to converge on collecting the same color resource to gain reward when they encounter each other and compare inventories. The winner's inventory is reset after an interaction. The matrix for the interaction is
\[ A_\text{row} = A_\text{col}^T  =
\begin{bmatrix}
1 & 0 & 0\\
0 & 1 & 0\\
0 & 0 & 1
\end{bmatrix}
\text{ .} \]

\textbf{Rationalizable Coordination in the Matrix}\footnote{For a video of \textit{Rationalizable Coordination in the Matrix}, see \url{https://youtu.be/BpHpoir06mY}.}
Same as Pure Coordination in the matrix except that differently colored resources are intrinsically of different values, suggesting an optimal color to converge on. The winner's inventory is reset after an interaction. The matrix for the interaction is
\[ A_\text{row} = A_\text{col}^T =
\begin{bmatrix}
1 & 0 & 0\\
0 & 2 & 0\\
0 & 0 & 3
\end{bmatrix}
\text{ .} \]

\textbf{Running with Scissors in the Matrix}\footnote{For a video of \textit{Running with Scissors in the Matrix}, see \url{https://youtu.be/oqYd4Ib5g70}.}
This environment first appeared in \citet{vezhnevets2020options}. Two individuals gather rock, paper or scissor resources in the environment and can challenge others to a 'rock, paper scissor' game, the outcome of which depends on the resources they collected.  It is possible---though not trivial---to observe the policy that one's partner is starting to implement and take countermeasures. This induces a wealth of possible feinting strategies.  Observation window is $5 \times 5$. Initial values for the inventory are $(1, 1, 1)$ instead of zeros.

\textbf{Stag Hunt in the Matrix}\footnote{For a video of \textit{Stag Hunt in the Matrix}, see \url{https://youtu.be/7fVHUH4siOQ}.}
Individuals collect resources that represent `hare' (red) or `stag' (green) and compare inventories in an encounter. Consequences of this inventory comparison are congruent with the classic Stag Hunt matrix game. This game exposes a tension between reward for the group and risk for the individual. The winner's inventory is reset after an interaction. The matrix for the interaction is
\[ A_\text{row} = A_\text{col}^T =
\begin{bmatrix}
4 & 0\\
2 & 2
\end{bmatrix}
\text{ .} \]

\textbf{Territory: Open}\footnote{For a video of \textit{Territory: Open}, see \url{https://youtu.be/3hB8lABa6nI}.}
Nine individuals can claim a territory for reward by coloring it. They can find a peaceful partition, but also have the option of irreversibly destroying a part of potentially rewarding territory rendering it useless for everyone. If one agent zaps another one then the zapped agent is removed from play until the end of the episode and all territory it claimed reverts to the neutral color.

\textbf{Territory: Rooms}\footnote{For a video of \textit{Territory: Rooms}, see \url{https://youtu.be/u0YOiShqzA4}.}
Same dynamics as \textit{Territory Open} except that individuals start in segregated rooms that strongly suggest a partition individuals could adhere to. They can break down the walls of these regions and invade each other's ``natural territory'', but the destroyed resources are lost forever. A peaceful partition is possible at the start of the episode, and the policy to achieve it is easy to implement. But if any agent gets too greedy and invades, it buys itself a chance of large rewards, but also chances inflicting significant chaos and deadweight loss on everyone if its actions spark wider conflict. The reason it can spiral out of control is that once an agent's neighbor has left their natural territory then it becomes rational to invade the space, leaving one's own territory undefended, creating more opportunity for mischief by others.

\section{Agent architecture details}

In the implementation of agents we aimed to stick with configurations proposed for the agents by their authors. We made sure that they use the same size convNets and LSTMs. We didn't perform any tuning of hyper-parameters and used the ones provided in original publications.

\paragraph{A3C:} The agent's network consists of a convNet with two layers with $16,32$ output channels, kernel shapes $8,4$, and strides $8,1$ respectively. It is followed by an MLP with two layers with 64 neurons each. All activation functions are ReLU. It is followed by an LSTM with $128$ units. Policy and baseline (for the critic) are produced by linear layers connected to the output of LSTM. 
We used an auxiliary loss~\cite{jaderberg2016reinforcement} for shaping the representation using contrastive predictive coding~\cite{oord2018representation}. CPC was discriminating between nearby time points via LSTM state representations (a standard augmentation in recent works with A3C). 
We used RMSProp optimizer with learning rate of $4*10^{-4}$, $\epsilon=10^{-5}$ and momentum set to zero and decay of $0.99$. Baseline cost $0.5$, entropy regularizer for policy at $0.003$. All agents used discount rate $\gamma=0.99$. 

\paragraph{V-MPO:} Has the same network as A3C with an extra PopArt~\cite{hessel2019multi} layer attached to the baseline for normalizing the gradients. The main difference is the use of target network and a different learning method. Initial $\eta=1, \alpha=5$. learning rate is set at $10^{-4}$.

\paragraph{OPRE:} Again, OPRE shares the basic structure of the network with A3C. The details of the architecture can be found in~\citet{vezhnevets2020options}. We used 16 options and a hierarchical critic implemented as an MLP with $64,64$ neurons. Critic has access to opponents observations. In~\citet{vezhnevets2020options} only one other player was present. To represent multiple coplayers, we simple concatenated the outputs of the visual stack (convNet+MLP) and their inventories (where appropriate) together, before feeding it into the critic.
OPRE had the same parameters for optimisation as AC3 plus two extra regularizers: i) entropy of policy over options set to $0.01$ and KL cost between critic and actor policies over options set at $0.01$. 

\paragraph{Puppet:} In sec. 6 we have described a way to construct a bot for a scenario using HRL in cases where the desired behavior of bots is complex and infeasible to train with a single pseudo-reward. 
The idea is to use environment events as goals for training basic policies and then combine them into complex ones using a pre-scripted high-level policy. The high-level policy is a set of ``if this event happens, activate that behaviour'' statements. The same set of environment events can be used to train basic policies and script the high-level one. 
We call this approach \emph{puppet}.

The puppet uses the same basic structure as other agents (ConvNet, MLP, LSTM), but has a hierarchical policy structure. The architecture is inspired by Feudal Networks~\cite{vezhnevets2017feudal}, but has several important differences. 
We represent goals as a one-hot vector $g$, which we embed into a continuous representation $e(g)$. We than feed $e$ as an extra input to the LSTM.
The network outputs several policies $\pi_z(a|x)$ and the final policy is a mixture $\pi(a|x)=\sum_z{\alpha(e) \pi_z(a|x)}$, where the mixture coefficients $\alpha(e)=\textbf{SoftMax}(e)$ are learnt from the embedding.
Notice, that instead of directly associating policies to goals, we allow the embedding to learn it through experience. 
To train the puppet to follow goals, we train it in the respective environment with goals switching at random intervals and rewarding the agent for following them.

\section{Training setup}

We use a distributed training set up similar to IMPALA (\citep{espeholt2018impala}), where the agent parameters are updated in a learner process on a GPU. Experience for learning is generated in an actor process. The actor handles inference and interactions with the environment on a CPU, using parameters cached periodically from the learner.

\section{Scenario details}

\subsection{Allelopathic Harvest}
\begin{SC}
    \item \emph{Focals are resident and a visitor prefers green.} This is a resident-mode scenario with a background population of A3C bots. The focal population must recognize that a single visitor bot has joined the game and is persisting in replanting berry patches in a color the resident population does not prefer (green instead of their preferred red). The focal agents should zap the visitor to prevent them from planting too much.
    
    \item \emph{Visiting a green preferring population.} This is a visitor-mode scenario with a background population of A3C bots. Four focal agents join twelve from the background population. In this case the background population strongly prefers green colored berries. They plant assiduously so it would be very difficult to stop them from making the whole map green. The four focal agents prefer red berries but need to recognize that in this case they need to give up on planting berries according to their preferred convention and instead join the resident population in following their dominant convention of green. Otherwise, if the focal agents persist in planting their preferred color then they can succeed only in making everyone worse off.
    
    \item [univ.] Agents face themselves in this setting. As a consequence, all players have congruent incentives alleviating the conflict between groups that desire different berry colors. This test also exposes free rider agents that rely on others to do the work of replanting.
\end{SC}

\subsection{Arena Running With Scissors in The Matrix}
\begin{SC}
    \item \emph{Versus gullible bots.} This is a half-and-half-mode scenario with a background population of A3C bots. Here the four focal agents interact in transient dyadic pairs with four bots from the background population that were trained to best respond to bots that implement pure rock, paper, or scissors strategies. They try to watch which resource their potential partners are collecting in order to pick the best response. However they are vulnerable to feinting strategies that trick them into picking the wrong counter.
    
    \item \emph{Versus mixture of pure bots.} This is a half-and-half-mode scenario with a background population of A3C bots. Here the four focal agents interact in transient dyadic pairs with four bots from the background population sampled from the full set of pure strategy bots. So some will be pure rock players, some pure paper, and some pure scissors. The task for the focal agents is to watch the other players, see what strategy they are implementing and act accordingly.
    
    \item \emph{Versus pure rock bots.} This is a half-and-half-mode scenario with a background population of A3C bots. Here the four focal agents interact in transient dyadic pairs with four bots from the background population that were trained with pseudorewards so they would implement pure rock strategies. The task of the focal agents is to collect paper resources, avoid one another, and target the rock players from the population to get high scores per interaction.
    
    \item \emph{Versus pure paper bots.} This is a half-and-half-mode scenario with a background population of A3C bots. Here the four focal agents interact in transient dyadic pairs with four bots from the background population that were trained with pseudorewards so they would implement pure paper strategies. The task of the focal agents is to collect scissors resources, avoid one another, and target the paper players from the population to get high scores per interaction.
    
    \item \emph{Versus pure scissors bots.} This is a half-and-half-mode scenario with a background population of A3C bots. Here the four focal agents interact in transient dyadic pairs with four bots from the background population that were trained with pseudorewards so they would implement pure scissors strategies. The task of the focal agents is to collect rock resources, avoid one another, and target the scissors players from the population to get high scores per interaction.
\end{SC}

\subsection{Bach or Stravinsky in The Matrix}

\begin{SC}
    \item \emph{Visiting pure Bach fans.} This is a visitor-mode scenario with a background population of A3C bots. Here the focal agent must work out that it is best to play Bach regardless of whether playing in a given episode as a row player or a column player. All potential interaction partners play Bach, so the the focal agent should follow that convention and thereby coordinate with them.
    
    \item \emph{Visiting pure Stravinsky fans.} This is a visitor-mode scenario with a background population of A3C bots. Here the focal agent must work out that it is best to play Stravinsky regardless of whether playing in a given episode as a row player or a column player. All potential interaction partners play Stravinsky, so the the focal agent should follow that convention and thereby coordinate with them.
    
    \item [univ.] This test may expose agents that have learned to be too stubborn.
\end{SC}

\subsection{Capture The Flag}

All bots in the background population trained with the following pseudoreward scheme: reward = 1 for zapping an avatar on the opposing team, reward = 2 for zapping the opposing team's flag carrier, reward = 3 for returning a flag dropped on the ground back to base by touching it, reward = 5 for picking up the opposing team's flag, and reward = 25 for capturing the opposing team's flag (or -25 when the opposing team does the same).

\begin{SC}
    \item \emph{Focal team versus shaped A3C bot team.} This is a half-and-half-mode scenario with a background population of A3C bots. Here a team of four focal agents square off against a team of four bots sampled from the background population. This is a purely competitive game so the goal is to defeat the opposing team. It requires teamwork and coordination to do so. Since focal players are always sampled from the same training population, they are familiar in this case with their teammates but the opposing team is unfamiliar to them (never encountered during training).
    
    \item \emph{Focal team versus shaped V-MPO bot team.} This is a half-and-half-mode scenario with a background population of V-MPO bots. Here a team of four focal agents square off against a team of four bots sampled from the background population. This is a purely competitive game so the goal is to defeat the opposing team. It requires teamwork and coordination to do so. Since focal players are always sampled from the same training population, they are familiar in this case with their teammates but the opposing team is unfamiliar to them (never encountered during training).
    
    \item \emph{Ad hoc teamwork with shaped A3C bots.} This is a visitor-mode scenario with a background population of A3C bots. It demands skills of ad-hoc teamwork. In this case the lone focal agent must coordinate with unfamiliar teammates in order to defeat a similarly unfamiliar opposing team.
    
    \item \emph{Ad hoc teamwork with shaped V-MPO bots.} This is a visitor-mode scenario with a background population of VMPO bots. It demands skills of ad-hoc teamwork. In this case the lone focal agent must coordinate with unfamiliar teammates in order to defeat a similarly unfamiliar opposing team.
\end{SC}

\subsection{Chemistry: Branched Chain Reaction}
The task features a reaction chain with two branch where the products from one can be fed back into the other to continue in a sustainable manner. While either branch can be run as long as there is material, running either more than the other runs out of some critical molecules. We train A3C both who are only rewarded for one branch. The evaluation agents can achieve their task if they manage to group with the right selection of other agents carrying suitable molecules, setting up a site where all the reaction keeps running sustainably.

\begin{SC}
    \item \emph{Focals meet X preferring bots.} This is a half-and-half-mode scenario with a background population of A3C bots specialized in branch $X$. The evaluated agents will have to find a way to combine with this particular kind of bots to set up desired chain reactions.
    
    \item \emph{Focals meet Y preferring bots.} This is a half-and-half-mode scenario with a background population of A3C bots specialized in branch $Y$.
    
    \item \emph{Focals are resident.} This is a resident-mode scenario where the evaluated agents are paired with a single A3C bot that will be a specialist in on of the two branches.
    
    \item \emph{Visiting another population.} This is a visitor-mode scenario with a background population of A3C bots where each is specialized on one of the cycles.
    
    \item [univ.] If policies have become so specialized that they are no longer able to function as generalists when necessary then they won't perform well in this test. Partner choice is especially important for this substrate which seems to make universalization scenarios somewhat easier.
\end{SC}

\subsection{Chemistry: Metabolic Cycles}
The task has some sub-cycles of reaction and we train bots who are are only rewarded for the food produced in one of the two, besides also getting rewarded for generating new energy. This training results in bots specialized in either cycle $A$ or cycle $B$ and that will ignore the other. In the first two tests it will be tested if the evaluated agents can adapts to being paired with bots specialized in one of the cycles and adapt to run the other.

\begin{SC}
    \item \emph{Focals meet A preferring bots.} This is a half-and-half-mode scenario with a background population of A3C bots specialized in cycle $A$. The evaluated agents will have to run the cycle $B$ side to be successful. 
    
    \item \emph{Focals meet B preferring bots.} This is a half-and-half-mode scenario with a background population of A3C bots specialized in cycle $B$. The evaluated agents will have to run the cycle $A$ side to be successful.
    
    \item \emph{Focals are resident.} This is a resident-mode scenario where the evaluated agents are paired with a single A3C bot that will be a specialist in on of the two cycles. For optimal returns the residents needs to make sure a good balance across tasks is achieved.
    
    \item \emph{Visiting another population.} This is a visitor-mode scenario with a background population of A3C bots where each is specialized on one of the cycles. For the most success the visitor needs to take on the most needed task that depends on the sampled bots.
    
    \item [univ.] This test exposes populations that learned specialization and division of labor strategies. If policies have become so specialized that they are no longer able to function as generalists when necessary then they won't perform well in this test.
\end{SC}

\subsection{Chicken in the Matrix}

\begin{SC}
    \item \emph{Meeting a mixture of pure bots.} This is a half-and-half-mode scenario with a background population of A3C bots. In this case the background population samples four bots uniformly over a set where each was pretrained to mostly play either hawk or dove. The focal agents must watch their interaction partners carefully to see if they have collected hawk and if they did, then either avoid them or play dove. If they want to play hawk then they should try to interact with other players who they saw collecting dove.
    
    \item \emph{Visiting a pure dove population.} This is a visitor-mode scenario where a single focal agent joins seven from the background population. All members of the background population were pretrained to specialize in playing dove. Thus the correct policy for the focal visitor is to play hawk.
    
    \item \emph{Focals are resident and visitors are hawks.} This is resident-mode scenario where three members of the background population join a focal resident population of five agents. In this case the background population bots specialize in playing hawk. The task for the resident agents is to seek always to interact in (hawk, dove) or (dove, dove) pairings, importantly avoiding (hawk, hawk).
    
    \item \emph{Visiting a gullible population.} This is a visitor-mode scenario where a single focal agent joins seven from the background population. In this case the background population bots trained alongside (mostly) pure hawk and dove agents, but were not themselves given any non-standard pseudoreward scheme. They learned to look for interaction partners who are collecting dove resources so they can defect on them by playing hawk. Thus they can be defeated by feinting to make them think you are playing dove when you actually will play hawk.
    
    \item \emph{Visiting grim reciprocators.} This is a visitor-mode scenario where two focal agents join a group of six from the background population. The background population bots are conditional cooperators. They play dove unless and until they are defected on by a partner (i.e.~ their partner chooses hawk). After that they will attempt to play hawk in all future encounters until the end of the episode. Such a strategy is often called ``Grim Trigger'' because it never forgives a defection (e.g.~\citet{axelrod2000six}). Note that they do not attempt to punish the specific player that defected on them in the first place, instead defecting indiscriminately on all future interaction partners, sparking a wave of recrimination that ends up causing all the background bots to defect after a short while. Since always last 1000 steps, the best strategy here for the focal agents would probably be to play dove up until near the end of the episode and then play hawk since no further retaliation will be possible at that point.
    
    \item [univ.] Dove-biased policies will perform relatively well but hawk-based policies will perform exceptionally badly here.
\end{SC}

\subsection{Clean Up}

\begin{SC}
    \item \emph{Visiting an altruistic population.} This is a visitor-mode scenario where three focal agents join four from the background population. The background bots are all interested primarily in cleaning the river. They rarely consume apples themselves. The right choice for agents of the focal population is to consume apples, letting the background population do the work of maintaining the public good.\label{cleanup_item:a}
    
    \item \emph{Focals are resident and visitors free ride.} This is a resident-mode scenario where four focal agents are joined by three from the background population. The background bots will only consume apples and never clean the river. It tests if the focal population is robust to having a substantial minority of the players (three out of seven) be ``defectors'' who never contribute to the public good. Focal populations that learned too brittle of a division of labor strategy, depending on the presence of specific focal players who do all the cleaning, are unlikely to do well in this scenario.\label{cleanup_item:b}

    \item \emph{Visiting a turn-taking population that cleans first.} This is a visitor-mode scenario where three focal agents join four agents from the background population. The background bots implement a strategy that alternates cleaning with eating every 250 steps. These agents clean in the first 250 steps. Focal agents that condition an unchangeable choice of whether to clean or eat based on what their coplayers do in the beginning of the episode (a reasonable strategy that self-play may learn) will perceive an opening to eat in this case. A better strategy would be to take turns with the background bots: clean when they eat and eat when they clean.\label{cleanup_item:c}

    \item \emph{Visiting a turn-taking population that eats first.} This is a visitor-mode scenario where three focal agents join four agents from the background population. As in \ref{cleanup_item:c}, the background bots implement a strategy that alternates cleaning with eating every 250 steps. However, in this case they start out by eating in the first 250 steps. If the focal population learned a policy resembling a ``grim trigger'' (cooperate until defected upon, then defect forever after), then such agents will immediately think they have been defected upon and retaliate. A much better strategy, as in SC2, is to alternate cleaning and eating out of phase with the background population.\label{cleanup_item:d} 

    \item \emph{Focals are visited by one reciprocator.} This is a resident-mode scenario where six focal agents are joined by one from the background population. In this case the background bot is a conditional cooperator. That is, it will clean the river as long as at least one other agent also cleans. Otherwise it will (try to) eat. Focal populations with at least one agent that reliably cleans at all times will be successful. It need not be the same cleaner all the time, and indeed the solution is more equal if all take turns.\label{cleanup_item:e}
    
    \item \emph{Focals are visited by two suspicious reciprocators.} This is resident-mode scenario where five focal agents are joined by two conditional cooperator agents from the background population. Unlike the conditional cooperator in \ref{cleanup_item:e}, these conditional cooperators have a more stringent condition: they will only clean if at least two other agents are cleaning. Otherwise they will eat. Thus, two cleaning agents from the focal population are needed at the outset in order to get the background bots to start cleaning. Once they have started then one of the two focal agents can stop cleaning. This is because the two background bots will continue to clean as long as, from each agent's perspective, there are at least two other agents cleaning. If any turn taking occurs among the focal population then they must be careful not to leave a temporal space when no focal agents are cleaning lest that cause the background bots to stop cooperating, after which they could only be induced to clean again by sending two focal agents to the river to clean\label{cleanup_item:f}
    
    \item \emph{Focals are visited by one suspicious reciprocator.} This is a resident-mode scenario where six focal agents are joined by one conditional cooperator agent from the background population. As in \ref{cleanup_item:f}, this conditional cooperator requires at least two other agents to be cleaning the river before it will join in and clean itself. The result is that the focal population must spare two agents to clean at all times, otherwise the background bot will not help. In that situation, the background bot joins as a third cleaner. But the dynamics of the substrate are such that it is usually more efficient if only two agents clean at once. Focal populations where agents have learned to monitor the number of other agents cleaning at any given time and return to the apple patch if more than two are present will have trouble in this scenario because once one of those agents leaves the river then so too will the background bot, dropping the total number cleaning from three down to one, which is even more suboptimal since a single agent cannot clean the river effectively enough to keep up a high rate of apple growth. The solution is that the focal population must notice the behavior of the conditional cooperating background bot and accept that there is no way to achieve the optimal number of cleaners (two), and instead opt for the least bad possibility where three agents (including the background conditional cooperator) all clean together.\label{cleanup_item:g}
    
    \item [univ.] This test exposes both free riders and overly altruistic policies. Both will get low scores here.
\end{SC}

\subsection{Collaborative Cooking: Impassable}

\begin{SC}
    \item \emph{Visiting a V-MPO population.} This is a visitor-mode scenario where one focal agent must join a group of three from the background population. The focal agent must observe the patterns of its coplayers' coordination and help out where ever it can. Above all, it must avoid getting in the way.
    
    \item \emph{Focals are resident.} This is a resident-mode scenario where three focal agents are joined by one from the background population. It tests if the coordination strategy learned by the focal population is robust to replacing one familiar agent with an unfamiliar agent from the background population, who might not be used to their conventions.
    
    \item [univ.] This test penalizes overspecialization.
\end{SC}

\subsection{Collaborative Cooking: Passable}

\begin{SC}
    \item \emph{Visiting uncoordinated generalists.} This is a visitor-mode scenario where one focal agent must join a group of three from the background population. The background population agents are all generalists who try to do all tasks themselves, and do not really coordinate much. The focal agent must recognize that the others are not being helpful and elect a similarly independent strategy.
    
    \item \emph{Focals are resident and visited by an uncoordinated generalist.} This is a resident-mode scenario where three focal agents are joined by one from the background population. The background bot is a generalist that attempts to do all tasks itself, and does not try to coordinate. The test is for the agents sampled from the focal population to find ways to cooperate despite the presence of this agent, who might frequently get in the way.
    
    \item [univ.] This test penalizes overspecialization.
\end{SC}

\subsection{Commons Harvest: Open}

\begin{SC}
    \item \emph{Focals are resident and visited by two zappers.} This is a resident-mode scenario where fourteen focal agents are joined by two from the background population. The two background bots are both especially interested in zapping, though not to the detriment of the rest of their policy. They do not act sustainably. The solution is to act sustainably while preventing the two background bots from depleting too many patches by harvesting them unsustainably. This can be done by zapping them whenever they approach a patch too closely.\label{commons_harvest_open_item_a}

    \item \emph{Focals are resident and visited by six zappers.} This is also a resident-mode scenario, but this time 10 focal agents are joined by six from the background community. The task is the same as  \ref{commons_harvest_open_item_a} but this time it is much harder because there are six unsustainable and zapping prone background bots instead of just two.
    
    \item [univ.] It is sometimes not problematic for a single individual to act greedily but if everyone acted that same way then it would be a catastrophe. This test exposes agents that learned to take advantage of such threshold effects.
    
\end{SC}

\subsection{Commons Harvest: Closed}

\begin{SC}
   \item \emph{Focals are resident and visited by two zappers.} This is a resident-mode scenario where fourteen focal agents are joined by two from the background population. Both background bots will act sustainably if given the chance by controlling the entrance to a room, though they are not perfectly consistent in this behavior. The solution is for as many individuals as possible to try to capture individual rooms by zapping any agent as they emerge from the corridor. This has the effect of making their harvesting behavior less interdependent. An agent that successfully controls the entrance to ``its territory'' is thus incentivized not to overharvest it. \label{commons_harvest_closed_item_a}
   
   \item \emph{Focals are resident and visited by six zappers.} Same as \ref{commons_harvest_closed_item_a} but harder since this time there are six background bots instead of just two. \label{commons_harvest_closed_item_b}
   
   \item \emph{Visiting a population of zappers.} This is a visitor-mode scenario where four focal agents join twelve background bots. It is similar to \ref{commons_harvest_closed_item_a} and \ref{commons_harvest_closed_item_b}. Despite the change from resident to visitor mode, the task for each individual focal agent is still more or less the same. The reason this is the case is that the goal behavior here is for agents to take actions to secure their independence. Once secured, an agent can control its own territory indefinitely, without any need to depend om others.
   
   \item [univ.] This test is similar to its analog in Commons Harvest Open.
\end{SC}

\subsection{Commons Harvest: Partnership}

This substrate raised the question: how can we make scenarios to test for abilities that state-of-the-art agents don't yet have?

Before starting this project we already knew that selfish RL agents acting in a group would learn to act unsustainably and cause a tragedy of the commons outcome in open field \textit{Commons Harvest} substrates like our \textit{Commons Harvest: Open} \cite{perolat2017, hughes2018inequity, mckee2020social}. We also already knew that agents that manage to get inside walled off regions could learn to zap invaders as they run down the corridor, and that after learning effective defense they would be reincentivized to act sustainably toward their resources since they could be sure to monopolize their future yield until the end of the episode, as in our \textit{Commons Harvest: Closed} substrate~\cite{perolat2017}. We wanted to create a test scenario where the walled regions each had two entrances so that pairs of agents would have to work together to defend ``their'' territory. Solving this test scenario should require agents to trust one another to (a) act sustainably with regard to their shared resources, and (b) competently guard one of the two entrances.

We knew from preliminary work that it is quite difficult to get current state-of-the-art agents to learn this kind of trust behavior. How then could we create bots to populate test scenarios that test for a behavior we don't know how to create in the first place? The solution we found was to create bots that would function as good partners for agents that truly got the concept of trust---even though they would not themselves truly ``understand'' it in a general way. We gave bots negative rewards during training for crossing invisible tiles located along the vertical midline of each room. Once agents came to regard those tiles as akin to walls, the game then resembled again the situation of \textit{Commons Harvest: Closed} where they reliably learn sustainable policies. A focal agent playing alongside one of these ``good partner'' bots would experience a coplayer that dutifully guards one or the other entrance and only ever collects apples from one side of the room. A focal agent capable of the limited form of trust demanded by this scenario should be able to cooperate with such a partner.

\begin{SC}
    \item \emph{Meeting good partners.} This is a half-and-half-mode scenario where eight focal agents join a background population consisting of eight good partner agents. The objective, for the agents lucky enough to reach the closed off regions first, is to collaborate with their partner to (a) defend the corridor an (b) act sustainably in harvesting their shared apple patches.\label{commons_harvest_partnership_a}
    
    \item \emph{Focals are resident and visitors are good partners.} This is a resident-mode scenario where twelve focal agents join four background good partner agents. It is similar to \ref{commons_harvest_partnership_a}. However, because there are more familiar individuals around (it is resident mode), the need for ad hoc cooperation with unfamiliar individuals is decreased.\label{commons_harvest_partnership_b}
    
    \item \emph{Visiting a population of good partners.} This is a visitor-mode scenario where four focal agents join twelve background good partner agents. Like \ref{commons_harvest_partnership_b}, this scenario differs from \ref{commons_harvest_partnership_a} in the extent to which success depends on ad hoc cooperation with unfamiliar individuals. In this case \emph{more} cooperation with unfamiliar individuals is required than in \ref{commons_harvest_partnership_a}.

    \item \emph{Focals are resident and visited by two zappers.} This is a resident-mode scenario where fourteen focal agents are joined by two background bots. Here the background bots are not good partners. They will frequently zap other players and act unsustainably whenever they get the chance to harvest. The solution is to partner with familiar agents, act sustainably, and cooperate to defend territories against invasion from the background bots.\label{commons_harvest_partnership_d}
    
    \item \emph{Focals are resident and visited by six zappers.} This is a resident-mode scenario where ten focal agents are joined by six background bots who are not good partners, act unsustainably, and zap frequently. It is a harder version of \ref{commons_harvest_partnership_d}. Since attempted invasions will occur more frequently, agents must be more skillful to prevent their territories from being overrun.
    
    \item \emph{Visiting a population of zappers.} This is a visitor-mode scenario there four focal agents are joined by twelve background bots who are not good partners, act unsustainably, and zap frequently. In this case it will usually be impossible for the small minority of focal agents to find one another and form an effective partnership. The best option from a menu with no good options is just to harvest apples as quickly as possible and act more-or-less unsustainably once invaded.
    
    \item [univ.] This test is similar to its analog in Commons Harvest Open.
\end{SC}

\subsection{King of the Hill}

\textit{King of the Hill} is an eight player game pitting two teams of four players each against one another.

The true (default) reward scheme is as follows. All members of a team get a reward of 1 on every frame that their team controls the hill region in the center of the map. If no team controls the map on a given frame (because no team controls more than 80\% of the hill) then no team gets a reward on that frame. No other events provide any reward. This reward scheme is always in operation at test time.

The alternative ``zap while in control'' reward scheme is as follows. Agents get a reward of 1 whenever both of the following conditions are satisfied simultaneously: (1) their team is in control of the hill, and (2) they just zapped a player of the opposing team, bringing their health to 0, and removing them from the game. This reward scheme was only used to train (some) background populations. Training with this scheme produces qualitatively different policies that still function effectively under the true reward scheme.

\begin{SC}
    \item \emph{Focal team versus default V-MPO bot team.} In this scenario a team composed entirely of focal agents must defeat a team composed entirely from a background population. This tests learned teamwork since a familiar team plays against an unfamiliar opposing team. In this case the background population was trained using the V-MPO algorithm, and used the substrate's true (default) reward scheme (see above). This background population tends to spend most of its time near the hill.
    
    \item \emph{Focal team versus shaped A3C bot team.} In this scenario a team composed entirely of focal agents must defeat a team composed entirely from a background population. This tests learned teamwork since a familiar team plays against an unfamiliar opposing team. In this case the background population was trained using the A3C algorithm, and used the `zap while in control' reward scheme (see above).
    
    \item \emph{Focal team versus shaped V-MPO bot team.} In this scenario a team composed entirely of focal agents must defeat a team composed entirely from a background population. This tests learned teamwork since a familiar team plays against an unfamiliar opposing team. In this case the background population was trained using the V-MPO algorithm, and used the `zap while in control' reward scheme. This causes the agents in the background population to implement a ``spawn camping'' policy. The counter-strategy is to evade opponents at the spawn location by running immediately toward the hill area to capture it, forcing the opposing team to abandon their position and return to defend the hill in a more chaotic and vulnerable fashion.
    
    \item \emph{Ad hoc teamwork with default V-MPO bots.} This scenario tests ad-hoc teamwork. One agent sampled from the focal population joins a team with three other agents from a background population to compete against another team of four agents sampled the same background population. If the focal agent works well with its unfamiliar teammates then it can tip the balance. In this case the background population was trained using the V-MPO algorithm and used the substrate's true (default) reward scheme. This background population tends to spend most of its time near the hill so in order to coordinate with them the focal agent should also spend time there. In particular, it should guard whichever entrance to the room containing the hill is least well guarded by its allies.

    \item \emph{Ad hoc teamwork with shaped A3C bots.} This scenario tests ad-hoc teamwork. One agent sampled from the focal population joins a team with three other agents from a background population to compete against another team of four agents sampled the same background population. If the focal agent works well with its unfamiliar teammates then it can tip the balance. In this case the background population was trained using the A3C algorithm and used the `zap while in control' reward scheme.
    
    \item \emph{Ad hoc teamwork with shaped V-MPO bots.} This scenario tests ad-hoc teamwork. One agent sampled from the focal population joins a team with three other agents from a background population to compete against another team of four agents sampled the same background population. If the focal agent works well with its unfamiliar teammates then it can tip the balance. In this case the background population was trained using the V-MPO algorithm and used the `zap while in control' reward scheme. To work well with these background bots who implement a ``spawn camping'' policy, the focal agent should follow them to the opposing team's base and help guard the escape routes by which spawning agents could get past them and take the hill. They must pick the side of the map to guard where less of their allies are stationed.
\end{SC}

\subsection{Prisoner's Dilemma in the Matrix}

Players are identifiable by color (randomized between episodes but maintained for each episode's duration).

\begin{SC}

    \item \emph{Visiting unconditional cooperators.} This is a visitor-mode scenario where one agent sampled from the focal population joins seven agents from a background population. In this case all the background bots will play cooperative strategies (mostly collecting `cooperate' resources and rarely collecting `defect'). The objective of a rational focal visitor is to exploit them by collecting `defect' resources.\label{prisoners_dilemma_itm_a}
    
    \item \emph{Focals are resident and visitors are unconditional cooperators.} This is a resident-mode scenario where six focal agents are joined by two agents from the background population. The background bots play cooperative strategies, as in \ref{prisoners_dilemma_itm_a}. A focal population will do well by defecting against these unfamiliar cooperators, but it should take care to identify them specifically. If they defect against all players regardless of identity then they will end up defecting on the focal population as well, lowering the overall score achieved.
    
    \item \emph{Focals are resident and visitors defect.} This is a resident-mode scenario where six focal agents are joined by two agents from the background population. The background bots play defecting strategies (mostly collecting `defect' resources). A focal population will do well by avoiding interaction with them and instead preferentially interacting with one another, and cooperating.\label{prisoners_dilemma_itm_c}
    
    \item \emph{Meeting gullible bots.} This is a half-and-half-mode scenario where four focal agents join four background bots. In this case the background population bots trained alongside (mostly) pure cooperator and defector bots, but were not themselves given any non-standard pseudoreward scheme. Thus they became bots that try to preferentially interact with partners who they have seen collecting cooperate resources. The best strategy against them is probably for the focal players to trick them into thinking they are defecting so they won't try to force interactions. This creates the necessary space to allow the focal population to preferentially interact (and cooperate) with one another.
    
    \item \emph{Visiting a population of grim reciprocators.} This is a visitor mode scenario where one focal agent joins seven from the background population. The background population bots are conditional cooperators. They will collect cooperate resources and attempt to play cooperate in all interactions until they have twice been defected upon by their partners. After that, they try to retaliate by collecting defect resources and aiming to defect in all remaining encounters till the end of the episode. Once they start defecting they do so indiscriminately. They do not try to specifically punish the players who defected on them. This sparks a wave of recrimination that ends up causing all the background bots to defect after a short while. Since episodes always last 1000 steps, the best strategy for the focal agent is probably to cooperate up until near the end of the episode and then defect once there is no remaining time for retaliation.\label{prisoners_dilemma_itm_generous_reciprocator}
    
    \item \emph{Visiting a population of hair-trigger grim reciprocators.} Just like \ref{prisoners_dilemma_itm_generous_reciprocator} but with background bot conditional cooperators operating on more of a hair trigger. That is, unlike \ref{prisoners_dilemma_itm_generous_reciprocator} where the bots will forgive one defection, the bots here will not forgive any defection at all. They begin defecting indiscriminately after the first time any agent defects on them.
    
    \item [univ.] Cooperation biased agents will do well while defection biased agents will do poorly here. If an agent learned a tit-for-tat style policy then it would also do well here.
\end{SC}

\subsection{Pure Coordination in the Matrix}

Unlike most other substrates, players here are not identifiable by color. All players look exactly the same as one another. It would be very difficult to track an individual for any significant length of time.

\begin{SC}

    \item \emph{Focals are resident and visitor is mixed.} This is a resident-mode scenario where seven focal agents are joined by one agent sampled from a background population. In this case the background population contains different kinds of specialist agents, each targeting one particular resource out of three. Focal agents need to watch other players to see what resource they are collecting and try to pick the same one. Most of the time they will be interacting with familiar others so whatever strategy they learned at test time will suffice. This scenario tests that their coordination is not disrupted by the presence of an unfamiliar other player who specializes in one particular resource.\label{pure_coordination_itm_a}

    \item \emph{Visiting resource A fans.} This is a visitor-mode scenario where one agent sampled from the focal population joins seven sampled from a background population. In this case the background population consists of specialists in resource A.\label{pure_coordination_itm_b}
    
    \item \emph{Visiting resource B fans.} Just like \ref{pure_coordination_itm_b} but with a background population consisting of specialists in resource B.
    
    \item \emph{Visiting resource C fans.} Just like \ref{pure_coordination_itm_b} but with a background population consisting of specialists in resource C.

    \item \emph{Meeting uncoordinated strangers.} This is a half-and-half-mode scenario where four focal agents join four background bots. As in \ref{pure_coordination_itm_a}, the background population contains players with all three different specializations. Thus they will normally be very uncoordinated. The objective for the focal population is to carefully watch which resources its potential partners will choose and select interaction partners on that basis, thus mainly interacting with familiar individuals who should all be playing the same strategy.
    
    \item [univ.] The main possible failure mode here is if too many agents try to collect too many of the same resources at one so they don't get distributed well and some agents miss the chance to collect one.
\end{SC}

\subsection{Rationalizable Coordination in the Matrix}

Unlike most other substrates, players here are not identifiable by color. All players look exactly the same as one another. It would be very difficult to track an individual for any significant length of time.

\begin{SC}

    \item \emph{Focals are resident and visitor is mixed.} This is a resident-mode scenario where seven focal agents are joined by one agent sampled from a background population. In this case the background population contains different kinds of specialist agents, each targeting one particular resource out of three. This scenario is similar to Pure Coordination in the Matrix \ref{prisoners_dilemma_itm_a} in that it tests that the focal population's coordination is not disrupted by the presence of an unfamiliar other player who specializes in one particular resource. The problem is more difficult here than in the pure coordination substrate though because all but one coordination choice is irrational. That is, while all choices are better than miscoordination, it is still clearly better to coordinate on some choices over others.\label{rationalizable_coordination_itm_a}

    \item \emph{Visiting resource A fans.} This is a visitor-mode scenario where one agent sampled from the focal population joins seven sampled from a background population. In this case the background population consists of specialists in resource A. It would be irrational for the group to coordinate on resource A since both resource B and C are better for everyone. \label{rationalizable_coordination_itm_b}
    
    \item \emph{Visiting resource B fans.} Just like \ref{rationalizable_coordination_itm_b} but with a background population consisting of specialists in resource B. Even though it is better to coordinate on resource B than on resource A, it is still irrational for the group to coordinate on it since resource C is strictly better.
    
    \item \emph{Visiting resource C fans.} Just like \ref{rationalizable_coordination_itm_b} but with a background population consisting of specialists in resource C. This is the resource that it is rational to coordinate on.
    
    \item \emph{Meeting uncoordinated strangers.} This is a half-and-half-mode scenario where four focal agents join four background bots. As in \ref{rationalizable_coordination_itm_a}, the background population contains players with all three different specializations. Thus they will normally be very uncoordinated. The objective for the focal population is to carefully watch which resources its potential partners will choose and select interaction partners on that basis, thus mainly interacting with familiar individuals who should all be collecting resource C since it is rational to do so.
    
    \item [univ.] This test is similar to its analog for Pure Coordination in the Matrix.
\end{SC}

\subsection{Running With Scissors in the Matrix}

This is a two-player zero-sum game. It was first introduced in \citet{vezhnevets2020options}.

\begin{SC}
    \item \emph{Versus gullible opponent.} Here the focal agent must defeat an opposing agent that was trained to best respond to agents playing pure strategies. The opponent should attempt to scout out what strategy the focal agent is playing so it can pick the appropriate counter. To defeat it, the focal agent should feint toward one resource and then collect the counter to its counter. So for example, if the focal agent successfully feinted that it would pick rock, inducing its opponent to pick paper, the focal agent should then collect and play scissors.\label{running_with_scissors_item:a} 
    
    \item \emph{Versus mixed strategy opponent.} Here the focal agent must defeat an opponent that was trained to play a relatively pure strategy: either rock, paper, or scissors. However, the specific opponent is sampled at test time so it could be any of those. To defeat it, the focal agent should scout out which pure strategy its opponent is playing and then collect the resources to implement its counter strategy.\label{running_with_scissors_item:b}
    
    \item \emph{Versus pure rock opponent.} Same as \ref{running_with_scissors_item:b}, but the opponent will always play rock.
    
    \item \emph{Versus pure paper opponent.} Same as \ref{running_with_scissors_item:b}, but the opponent will always play paper.
    
    \item \emph{Versus pure scissors opponent.} Same as \ref{running_with_scissors_item:b}, but the opponent will always play scissors.
\end{SC}

\subsection{Stag Hunt in the Matrix}

\begin{SC}

    \item \emph{Visiting a population of stags.} This is a visitor-mode scenario where one agent sampled from the focal population joins seven sampled from a background population. All background bots have been trained to primarily play stag. The right solution is for the focal visitor agent to also play stag.\label{stag_hunt_itm_a}
    
    \item \emph{Visiting a population of hares.} This is a visitor-mode scenario where one agent sampled from the focal population joins seven sampled from a background population. This test is just like \ref{stag_hunt_itm_a} except the background population was trained primarily to play hare. The right solution is for the focal visitor agent to also play hare.
    
    \item \emph{Visiting a population of grim reciprocators.} This is a visitor-mode scenario where two focal agents join six bots from the background population. The background population bots are reciprocators. They start out by playing stag but if any of their interaction partners plays hare then they become triggered to play hare themselves in all their future interactions till the end of the episode.
    
    \item [univ.] Stag biased agents will do better than hare biased agents as long as they can avoid running out of stag resources due to individuals picking up more than they need to commit to the strategy.
\end{SC}

\subsection{Territory: Open}

\begin{SC}

    \item \emph{Focals are resident and visited by a shaped bot.} This is a resident-mode scenario where eight agents from the focal population are joined by one agent sampled from a background population. The background bot is typically quite active, it runs around the map and claims as much territory as it can. It doesn't use its zapping beam very often. The right response to it is often to zap it early in the episode so it can't get in the way of the rest of the agents fairly dividing up the resources according to whatever convention they agree on (i.e. whatever convention emerged during training).\label{territory_open_a}
    
    \item \emph{Visiting a population of shaped bots.} This is a visitor-mode scenario where one focal agent joins eight from a background population. The background bots all behave as described in \ref{territory_open_a}. They color a lot of territory but frequently color over one another's territory. The focal visitor could follow their conventional behavior and run around the map with them claiming and reclaiming as they go. Or alternatively, it may be possible to defend an area of the map and zap them if they approach too close. The latter strategy would likely achieve higher rewards if successful.
    
    \item [univ.] This test exposes overly aggressive policies. They will perform very poorly.
\end{SC}

\subsection{Territory: Rooms}

\begin{SC}

    \item \emph{Focals are resident and visited by an aggressor.} This is a resident-mode scenario where eight agents from the focal population are joined by one agent sampled from a background population. The background population consists of agents that, although they did not themselves experience any special reward shaping, they trained alongside other agents that were incentivized to be extremely aggressive and destructive in their zapping behavior. As a result, they are fairly aggressive themselves while still being interested primarily in claiming territory. A focal population where agents have not forgotten how to defend themselves from attack will do much better in this scenario than a focal population that implements the same fair division of the resources but forgot self-defense.\label{territory_rooms_a}
    
    \item \emph{Visiting a population of aggressors.} This is a visitor-mode scenario where one focal agent joins eight from a background population. The background population is the same as in \ref{territory_rooms_a}. The focal visitor agent must be forever on its guard since neighboring agents may attack their territory at any time, permanently destroying resources on the way in. Since the larger set of background bots are so likely to fight among themselves, the optimal strategy for the focal visitor agent is often to invade after a battle and claim extra territory in areas where other agents have been zapped and removed.
    
    \item [univ.] This test exposes overly aggressive policies. They will perform very poorly.
\end{SC}

\onecolumn
\section{Raw performance scores}
\begin{longtable}{ll|rrrrrr|rr}
\caption{Focal per-capita returns.}\\
\toprule
                 & agent &                       A3C &                    A3C PS &                      OPRE &                   OPRE PS &                     V-MPO &                  V-MPO PS &                 exploiter &                    random \\
substrate & scenario &                           &                           &                           &                           &                           &                           &                           &                           \\
\midrule
\endhead
\midrule
\multicolumn{10}{r}{{Continued on next page}} \\
\midrule
\endfoot

\bottomrule
\endlastfoot
\multirow{4}{3cm}{Allelopathic Harvest} & S-P &                      63.8 &                      60.4 &                      61.4 &                      92.4 &                      33.1 &                      48.5 &                       n/a &                     -17.8 \\
                 & SC 0 &                      42.6 &                      37.9 &                      36.4 &                      50.6 &                      60.7 &                      36.2 &                      66.8 &                     -18.9 \\
                 & SC 1 &                     137.5 &                      14.0 &                     140.3 &                      39.1 &                     125.5 &                      32.8 &                     161.2 &                       1.5 \\
                 & univ. &                      51.1 &                      41.1 &                      43.2 &                      74.9 &                      32.9 &                      42.7 &                       n/a &                     -18.1 \\
\cline{1-10}
\multirow{6}{3cm}{Arena Running with Scissors in the Matrix} & S-P &                      0.00 &                     -0.00 &                     -0.00 &                       n/a &                      0.00 &                       n/a &                       n/a &                      0.00 \\
                 & SC 0 &                     -0.01 &                     -0.01 &                     -0.00 &                       n/a &                      0.03 &                       n/a &                      0.11 &                     -0.01 \\
                 & SC 1 &                     -0.00 &                     -0.00 &                      0.00 &                       n/a &                     -0.02 &                       n/a &                      0.12 &                     -0.01 \\
                 & SC 2 &                      0.00 &                     -0.00 &                      0.01 &                       n/a &                      0.06 &                       n/a &                      0.60 &                      0.00 \\
                 & SC 3 &                      0.01 &                      0.01 &                      0.00 &                       n/a &                      0.03 &                       n/a &                      0.56 &                      0.01 \\
                 & SC 4 &                      0.01 &                      0.01 &                      0.01 &                       n/a &                      0.09 &                       n/a &                      0.66 &                      0.00 \\
\cline{1-10}
\multirow{4}{3cm}{Bach or Stravinsky in the Matrix} & S-P &                      10.8 &                      10.2 &                      10.4 &                      10.6 &                       7.8 &                       8.1 &                       n/a &                       0.4 \\
                 & SC 0 &                       8.3 &                       1.9 &                       8.9 &                       1.6 &                       1.9 &                       0.6 &                      14.9 &                       0.1 \\
                 & SC 1 &                       1.7 &                       0.9 &                       1.3 &                       1.0 &                       2.3 &                       2.2 &                       6.5 &                       0.4 \\
                 & univ. &                      10.6 &                       8.8 &                      10.1 &                       9.6 &                       6.1 &                       7.2 &                       n/a &                       0.4 \\
\cline{1-10}
\multirow{5}{3cm}{Capture the Flag} & S-P &                       0.0 &                       0.0 &                       0.0 &                       n/a &                       0.0 &                       n/a &                       n/a &                       0.0 \\
                 & SC 0 &                     -12.8 &                     -12.8 &                     -12.8 &                       n/a &                     -12.3 &                       n/a &                      -1.8 &                     -12.9 \\
                 & SC 1 &                     -14.4 &                     -14.1 &                     -14.3 &                       n/a &                     -14.6 &                       n/a &                      -2.6 &                     -14.2 \\
                 & SC 2 &                      -3.8 &                      -3.7 &                      -3.7 &                       n/a &                      -4.1 &                       n/a &                       5.7 &                      -3.8 \\
                 & SC 3 &                      -3.8 &                      -3.6 &                      -3.6 &                       n/a &                      -4.2 &                       n/a &                       5.5 &                      -3.6 \\
\cline{1-10}
\multirow{6}{3cm}{Chemistry: Branched Chain Reaction} & S-P &                      30.6 &                       9.3 &                       5.1 &                      24.9 &                     177.6 &                      84.4 &                       n/a &                       0.0 \\
                 & SC 0 &                      24.5 &                       8.5 &                       4.1 &                       0.9 &                     100.8 &                      10.7 &                      85.0 &                       0.0 \\
                 & SC 1 &                      12.5 &                       6.9 &                       4.6 &                       0.3 &                      81.8 &                      15.1 &                      59.4 &                       0.0 \\
                 & SC 2 &                      18.3 &                      10.3 &                       4.8 &                       2.7 &                     127.8 &                      20.4 &                      91.1 &                       0.1 \\
                 & SC 3 &                      19.8 &                       1.1 &                       3.8 &                       0.5 &                      56.5 &                       9.5 &                      20.4 &                       0.0 \\
                 & univ. &                       6.4 &                       0.9 &                       3.8 &                       0.2 &                     138.0 &                       3.1 &                       n/a &                       0.0 \\
\cline{1-10}
\multirow{6}{3cm}{Chemistry: Metabolic Cycles} & S-P &                      82.3 &                      49.1 &                     237.9 &                       5.1 &                     185.6 &                      73.7 &                       n/a &                       0.3 \\
                 & SC 0 &                      62.5 &                       8.9 &                     165.7 &                       3.1 &                     118.0 &                      40.5 &                     206.1 &                       0.6 \\
                 & SC 1 &                      31.1 &                      26.9 &                     139.1 &                       5.8 &                      96.0 &                      22.0 &                     162.3 &                       0.6 \\
                 & SC 2 &                      55.9 &                      19.7 &                     177.6 &                       4.5 &                     122.7 &                      41.4 &                     153.8 &                       0.4 \\
                 & SC 3 &                      69.8 &                      16.6 &                     138.6 &                       9.4 &                     148.2 &                      26.3 &                     156.4 &                       3.1 \\
                 & univ. &                       0.5 &                       0.4 &                       1.5 &                       0.9 &                       6.9 &                       0.4 &                       n/a &                       0.3 \\
\cline{1-10}
\multirow{7}{3cm}{Chicken in the Matrix} & S-P &                      15.6 &                      23.8 &                      14.8 &                      24.7 &                      12.5 &                      22.3 &                       n/a &                       1.0 \\
                 & SC 0 &                      15.6 &                      12.0 &                      14.1 &                      11.6 &                      12.0 &                      12.3 &                      15.2 &                       1.2 \\
                 & SC 1 &                      70.2 &                      26.4 &                      72.4 &                      21.2 &                      76.7 &                      33.7 &                      98.9 &                       3.9 \\
                 & SC 2 &                      11.9 &                      11.1 &                      10.9 &                      11.2 &                       8.4 &                      10.8 &                      14.1 &                       0.8 \\
                 & SC 3 &                      14.1 &                       7.7 &                      13.2 &                       7.0 &                      13.9 &                       7.6 &                      14.3 &                       0.7 \\
                 & SC 4 &                      30.1 &                      15.6 &                      26.2 &                      15.6 &                      24.6 &                      17.7 &                      36.4 &                       4.9 \\
                 & univ. &                      14.8 &                      21.1 &                      14.6 &                      21.7 &                      11.5 &                      21.5 &                       n/a &                       1.0 \\
\cline{1-10}
\multirow{9}{3cm}{Clean Up} & S-P &                       0.0 &                      36.9 &                       0.0 &                      72.8 &                       0.0 &                     188.6 &                       n/a &                       0.1 \\
                 & SC 0 &                      91.5 &                     380.3 &                     330.6 &                     290.7 &                     280.5 &                     354.5 &                     722.6 &                      69.1 \\
                 & SC 1 &                       0.0 &                       3.5 &                       0.0 &                      18.7 &                       0.0 &                      28.4 &                       0.0 &                       0.0 \\
                 & SC 2 &                      39.2 &                     156.8 &                     181.3 &                     122.5 &                     137.7 &                     185.1 &                     385.9 &                      32.3 \\
                 & SC 3 &                      28.8 &                     113.6 &                     126.0 &                      90.1 &                      93.9 &                     159.3 &                     225.6 &                      23.8 \\
                 & SC 4 &                      35.7 &                     144.9 &                      76.8 &                     124.9 &                      65.8 &                     250.5 &                     160.3 &                      29.0 \\
                 & SC 5 &                      53.5 &                     194.4 &                     134.4 &                     150.6 &                      81.8 &                     271.7 &                     256.1 &                      42.0 \\
                 & SC 6 &                      21.0 &                     114.6 &                      34.0 &                      93.2 &                      21.3 &                     224.7 &                     126.1 &                      18.4 \\
                 & univ. &                       0.0 &                      20.9 &                       0.0 &                      16.7 &                       0.9 &                      13.5 &                       n/a &                       0.0 \\
\cline{1-10}
\multirow{4}{3cm}{Collaborative Cooking: Impassable} & S-P &                       0.0 &                       0.0 &                       0.0 &                       0.0 &                       0.0 &                       0.0 &                       n/a &                       0.0 \\
                 & SC 0 &                     211.7 &                     203.0 &                     193.0 &                     192.6 &                     125.0 &                     165.2 &                     268.9 &                     198.6 \\
                 & SC 1 &                       0.2 &                       0.3 &                       0.2 &                       0.2 &                       0.0 &                       0.2 &                      34.0 &                       0.0 \\
                 & univ. &                       0.0 &                       0.0 &                       0.0 &                       0.0 &                       0.0 &                       0.0 &                       n/a &                       0.0 \\
\cline{1-10}
\multirow{4}{3cm}{Collaborative Cooking: Passable} & S-P &                       0.0 &                       0.0 &                       0.0 &                       0.0 &                       0.0 &                       0.0 &                       n/a &                       0.0 \\
                 & SC 0 &                     252.2 &                     252.7 &                     230.6 &                     206.2 &                     161.5 &                     251.7 &                     268.0 &                     247.8 \\
                 & SC 1 &                     108.9 &                     108.9 &                     107.5 &                      76.2 &                      67.3 &                     108.8 &                     118.3 &                     108.8 \\
                 & univ. &                       0.0 &                       0.0 &                       0.0 &                       0.0 &                       0.0 &                       0.0 &                       n/a &                       0.0 \\
\cline{1-10}
\multirow{5}{3cm}{Commons Harvest: Closed} & S-P &                      50.6 &                      16.8 &                      52.9 &                      27.6 &                      99.4 &                      26.1 &                       n/a &                       0.3 \\
                 & SC 0 &                      48.3 &                      13.3 &                      47.6 &                      28.6 &                      96.8 &                      26.0 &                      36.1 &                       0.2 \\
                 & SC 1 &                      42.5 &                       8.0 &                      37.0 &                      19.2 &                      91.4 &                      16.3 &                      29.3 &                       0.0 \\
                 & SC 2 &                      44.0 &                       1.5 &                      34.4 &                       4.7 &                      81.2 &                       4.7 &                       0.0 &                       0.0 \\
                 & univ. &                      29.0 &                       5.8 &                      30.0 &                      17.3 &                      78.5 &                       8.9 &                       n/a &                       0.3 \\
\cline{1-10}
\multirow{4}{3cm}{Commons Harvest: Open} & S-P &                      16.1 &                      51.2 &                      16.2 &                      64.0 &                      16.3 &                      65.7 &                       n/a &                      30.0 \\
                 & SC 0 &                      16.1 &                      16.4 &                      16.1 &                      23.7 &                      16.2 &                      24.5 &                      16.1 &                      10.1 \\
                 & SC 1 &                      16.7 &                       5.5 &                      16.7 &                      10.4 &                      15.9 &                      13.4 &                      16.5 &                       2.3 \\
                 & univ. &                      16.1 &                      51.1 &                      16.2 &                      43.7 &                      16.3 &                      65.9 &                       n/a &                      30.3 \\
\cline{1-10}
\multirow{8}{3cm}{Commons Harvest: Partnership} & S-P &                      19.5 &                      20.1 &                      19.0 &                      39.5 &                      21.1 &                      45.3 &                       n/a &                       0.2 \\
                 & SC 0 &                      17.7 &                       1.6 &                      17.3 &                       8.7 &                      28.3 &                      12.5 &                      24.7 &                       0.0 \\
                 & SC 1 &                      19.0 &                       8.3 &                      18.8 &                      21.2 &                      24.6 &                      25.6 &                      19.0 &                       0.2 \\
                 & SC 2 &                      17.3 &                       0.9 &                      17.5 &                       3.6 &                      35.1 &                       4.6 &                      44.0 &                       0.0 \\
                 & SC 3 &                      23.9 &                      18.7 &                      22.9 &                      39.1 &                      24.8 &                      49.5 &                      20.5 &                       0.3 \\
                 & SC 4 &                      36.2 &                      21.0 &                      36.0 &                      47.6 &                      38.7 &                      52.1 &                      33.7 &                       0.1 \\
                 & SC 5 &                      71.6 &                      21.8 &                      71.5 &                      45.9 &                     102.4 &                      52.6 &                     224.4 &                       0.1 \\
                 & univ. &                      13.0 &                       6.4 &                      11.0 &                      10.9 &                      21.9 &                      12.6 &                       n/a &                       0.3 \\
\cline{1-10}
\multirow{7}{3cm}{King of the Hill} & S-P &                       0.0 &                       0.0 &                       0.0 &                       n/a &                       0.0 &                       n/a &                       n/a &                       0.0 \\
                 & SC 0 &                    -969.5 &                    -990.4 &                    -978.2 &                       n/a &                      -3.2 &                       n/a &                    -990.4 &                    -990.4 \\
                 & SC 1 &                    -895.0 &                    -990.0 &                    -942.4 &                       n/a &                     627.8 &                       n/a &                    -991.1 &                    -990.6 \\
                 & SC 2 &                    -976.5 &                    -990.4 &                    -987.6 &                       n/a &                    -155.9 &                       n/a &                    -990.6 &                    -990.6 \\
                 & SC 3 &                    -535.4 &                    -600.9 &                    -547.7 &                       n/a &                     -55.4 &                       n/a &                    -607.9 &                    -603.8 \\
                 & SC 4 &                     -44.6 &                     -84.5 &                     -12.2 &                       n/a &                     402.4 &                       n/a &                    -105.6 &                    -144.8 \\
                 & SC 5 &                    -598.7 &                    -707.8 &                    -630.7 &                       n/a &                     -38.9 &                       n/a &                    -661.4 &                    -652.5 \\
\cline{1-10}
\multirow{8}{3cm}{Prisoners Dilemma in the Matrix} & S-P &                       7.1 &                      20.0 &                       7.1 &                      22.8 &                       7.3 &                      18.6 &                       n/a &                       0.9 \\
                 & SC 0 &                      30.9 &                       5.7 &                      28.2 &                       5.5 &                      21.3 &                       6.0 &                      55.7 &                       1.9 \\
                 & SC 1 &                      10.7 &                      13.7 &                      10.8 &                      11.9 &                      10.8 &                      13.8 &                      11.5 &                       1.8 \\
                 & SC 2 &                       7.4 &                       6.3 &                       7.3 &                       6.6 &                       7.0 &                       6.2 &                       7.8 &                       0.5 \\
                 & SC 3 &                       8.2 &                       3.2 &                       8.0 &                       3.3 &                       7.4 &                       3.4 &                      10.1 &                       0.4 \\
                 & SC 4 &                      52.3 &                       8.7 &                      46.8 &                       7.5 &                      33.3 &                      10.1 &                      60.8 &                       5.0 \\
                 & SC 5 &                      32.9 &                       9.1 &                      30.5 &                       7.9 &                      24.1 &                       9.1 &                      36.8 &                       5.4 \\
                 & univ. &                       7.1 &                      19.1 &                       6.9 &                      21.9 &                       7.2 &                      18.3 &                       n/a &                       0.9 \\
\cline{1-10}
\multirow{7}{3cm}{Pure Coordination in the Matrix} & S-P &                       4.4 &                       3.9 &                       4.5 &                       4.2 &                       4.1 &                       3.0 &                       n/a &                       0.1 \\
                 & SC 0 &                       4.0 &                       3.3 &                       4.1 &                       3.8 &                       3.9 &                       2.8 &                       4.4 &                       0.1 \\
                 & SC 1 &                       2.7 &                       0.8 &                       1.1 &                       0.8 &                       1.8 &                       1.1 &                       3.7 &                       0.2 \\
                 & SC 2 &                       1.1 &                       0.7 &                       0.9 &                       0.4 &                       1.0 &                       0.8 &                       3.2 &                       0.1 \\
                 & SC 3 &                       1.6 &                       0.4 &                       1.0 &                       0.8 &                       1.5 &                       0.8 &                       3.2 &                       0.2 \\
                 & SC 4 &                       3.6 &                       2.6 &                       3.6 &                       3.0 &                       3.5 &                       2.4 &                       4.0 &                       0.2 \\
                 & univ. &                       4.3 &                       3.0 &                       4.4 &                       4.1 &                       4.0 &                       2.7 &                       n/a &                       0.1 \\
\cline{1-10}
\multirow{7}{3cm}{Rationalizable Coordination in the Matrix} & S-P &                      12.6 &                      11.2 &                      12.5 &                      12.6 &                      10.9 &                       7.1 &                       n/a &                       0.2 \\
                 & SC 0 &                      11.4 &                      10.1 &                      11.3 &                      11.6 &                      10.2 &                       7.0 &                      11.9 &                       0.3 \\
                 & SC 1 &                       3.1 &                       2.1 &                       3.3 &                       2.6 &                       2.0 &                       1.7 &                       7.7 &                       0.3 \\
                 & SC 2 &                       1.3 &                       0.5 &                       1.3 &                       0.4 &                       0.9 &                       0.6 &                       6.4 &                       0.2 \\
                 & SC 3 &                       6.0 &                       5.3 &                       5.6 &                       5.7 &                       5.6 &                       4.2 &                       8.7 &                       0.4 \\
                 & SC 4 &                      10.4 &                       8.6 &                      10.2 &                      10.2 &                       9.4 &                       6.5 &                      13.1 &                       0.4 \\
                 & univ. &                      12.6 &                       9.3 &                      10.8 &                      12.3 &                      10.5 &                       6.4 &                       n/a &                       0.2 \\
\cline{1-10}
\multirow{6}{3cm}{Running with Scissors in the Matrix} & S-P &                      0.00 &                      0.00 &                      0.00 &                       n/a &                      0.00 &                       n/a &                       n/a &                      0.00 \\
                 & SC 0 &                     -0.00 &                      0.00 &                      0.02 &                       n/a &                      0.01 &                       n/a &                      0.00 &                     -0.00 \\
                 & SC 1 &                     -0.01 &                     -0.01 &                      0.06 &                       n/a &                      0.00 &                       n/a &                      0.00 &                     -0.02 \\
                 & SC 2 &                     -0.01 &                     -0.01 &                      0.07 &                       n/a &                      0.01 &                       n/a &                      0.39 &                     -0.01 \\
                 & SC 3 &                     -0.00 &                     -0.00 &                      0.03 &                       n/a &                      0.00 &                       n/a &                      0.38 &                     -0.01 \\
                 & SC 4 &                     -0.02 &                     -0.02 &                      0.05 &                       n/a &                      0.00 &                       n/a &                      0.37 &                     -0.02 \\
\cline{1-10}
\multirow{5}{3cm}{Stag Hunt in the Matrix} & S-P &                      29.8 &                      26.0 &                      29.5 &                      41.6 &                      26.8 &                      25.4 &                       n/a &                       1.2 \\
                 & SC 0 &                      34.2 &                      20.9 &                      34.1 &                      16.4 &                      29.8 &                      18.1 &                      65.9 &                       2.7 \\
                 & SC 1 &                      28.6 &                       6.6 &                      26.4 &                       2.4 &                      20.8 &                       3.7 &                      54.4 &                       1.3 \\
                 & SC 2 &                      39.3 &                      24.0 &                      40.4 &                      27.7 &                      33.0 &                      25.7 &                      53.8 &                       4.0 \\
                 & univ. &                      28.5 &                      19.1 &                      27.2 &                      32.1 &                      25.2 &                      23.5 &                       n/a &                       1.3 \\
\cline{1-10}
\multirow{4}{3cm}{Territory: Open} & S-P &                      77.3 &                      19.7 &                      60.6 &                      58.5 &                      62.6 &                      23.0 &                       n/a &                       4.8 \\
                 & SC 0 &                      53.1 &                       7.3 &                      44.3 &                      21.1 &                      40.5 &                      17.1 &                      45.1 &                       4.7 \\
                 & SC 1 &                      60.4 &                       7.1 &                      58.2 &                      19.3 &                      94.2 &                      39.0 &                      60.5 &                       7.5 \\
                 & univ. &                      63.0 &                      14.9 &                      54.1 &                      47.0 &                      57.7 &                      25.6 &                       n/a &                       4.9 \\
\cline{1-10}
\multirow{4}{3cm}{Territory: Rooms} & S-P &                     236.3 &                     199.4 &                     220.3 &                     112.2 &                      54.4 &                     233.3 &                       n/a &                      10.4 \\
                 & SC 0 &                     186.4 &                     160.4 &                     158.0 &                      70.6 &                     128.0 &                     161.5 &                     203.8 &                       9.1 \\
                 & SC 1 &                      27.3 &                      12.9 &                       6.2 &                      15.3 &                       5.7 &                       4.3 &                     273.4 &                       1.8 \\
                 & univ. &                     230.2 &                     189.8 &                     192.6 &                     134.6 &                      56.0 &                     220.6 &                       n/a &                      10.2 \\
\end{longtable}
\clearpage
\twocolumn

\section{Acknowledgements}

The authors would like to thank: Richard Everett, DJ Strouse, Kevin McKee, and Edward Hughes for contributing to substrate development; Mary Cassin for contributing artwork; and Steven Wheelwright for help building an early version of the evaluation system.

\section{References}
\vspace{-5mm}

\renewcommand{\refname}{}
\bibliography{biblio}

\end{document}